\documentclass[twocolumn,10pt,aps,pra,showpacs,superscriptaddress,nobalancelastpage,longbibliography,nofootinbib]{revtex4-2}

\usepackage{graphicx}
\usepackage{xcolor}
\usepackage{braket}
\usepackage[colorlinks]{hyperref}
\usepackage{rotating}
\usepackage{svg}
\usepackage{booktabs}
\usepackage[caption=false]{subfig}
\usepackage{multirow}
\usepackage{amsmath}

\hypersetup{
    colorlinks=true,
    linkcolor=blue,
    filecolor=magenta,
    urlcolor=blue,
}

\newcommand{\beginsupplement}{
  \setcounter{table}{0}  
  \renewcommand{\thetable}{S\arabic{table}} 
  \setcounter{figure}{0} 
  \renewcommand{\thefigure}{S\arabic{figure}}
  \setcounter{section}{0}
  \setcounter{equation}{0}
  \renewcommand{\theequation}{S\arabic{equation}}
}

\begin{document}

\title{LUCI on IBM Hardware: Error Suppression with Almost Half Syndrome Density}

\author{Younghun Kim}
\email{younghunk@student.unimelb.edu.au}
\affiliation{School of Physics, The University of Melbourne, Parkville, 3010, Victoria, Australia}
\affiliation{Data61, CSIRO, Clayton, 3168, Victoria, Australia}

\author{Spiro Gicev}
\affiliation{School of Physics, The University of Melbourne, Parkville, 3010, Victoria, Australia}

\author{Martin Sevior}
\affiliation{School of Physics, The University of Melbourne, Parkville, 3010, Victoria, Australia}

\author{Muhammad Usman}
\affiliation{School of Physics, The University of Melbourne, Parkville, 3010, Victoria, Australia}
\affiliation{Data61, CSIRO, Clayton, 3168, Victoria, Australia}

\begin{abstract}
Long-lived logical qubits are essential for fault-tolerant quantum computation. However, the practical performance of traditional error correction protocols relies on performing specific syndrome circuits, causing vulnerability to hardware defects and imposing rigid connectivity constraints. Recent theoretical findings have proposed that flexible subroutine circuits within the LUCI framework can maintain space-time distance in the presence of isolated or broken components, albeit at the expense of temporal distance. However, these approaches have solely targeted defect avoidance and have not yet been demonstrated to suppress errors with reduced temporal distances on physical hardware. In this work, we propose a reset-free scenario for the LUCI framework and experimentally benchmark it on IBM quantum hardware. By asymmetrically scaling the $X$ or $Z$ distance, we compare our reset-free approach against the standard surface code and successfully demonstrate error suppression ratios for targeted logical Pauli errors. Remarkably, despite a nearly halved syndrome density in time, which requires two subroutine rounds for full syndrome extraction, the LUCI framework remains competitive with the rotated surface code implementation. In the LUCI framework, we observe error suppression of $1.75(10)$ for logical $X$ errors and $1.93(12)$ for logical $Z$ errors, whereas the standard approach yields $ 1.58(13)$ and $2.44(7)$, respectively. These results demonstrate that dynamic codes outperform standard methods by avoiding highly noisy components, even without physical defects, while preserving logical boundaries. Our findings challenge the conventional dependency on static fault-tolerant architectures by verifying the feasibility and efficacy of the LUCI framework on physical hardware and pave the way for hybrid, hardware-compatible code designs in quantum computing.
\end{abstract}

\maketitle

\section{INTRODUCTION}
Scalable quantum computation requires logical qubits capable of preserving their state over execution times sufficient to achieve a practical computation regime~\cite{yoder_tour_2025,gidney_how_2025,webster_pinnacle_2026,cain_shors_2026,tripier_fault-tolerant_2026}. Quantum Error Correction (QEC) could enable the realization of such logical qubits even when implemented on fragile qubits of the Noisy Intermediate-Scale Quantum (NISQ)~\cite{shor_scheme_1995,preskill_quantum_2018}. Among competing QEC protocols within the Calderbank-Shor-Steane code family~\cite{shor_scheme_1995,calderbank_good_1996,steane_error_1996}, the surface code family is most practical for locally constrained processors due to its high threshold values~\cite{bravyi_quantum_1998,dennis_topological_2002,kitaev_fault-tolerant_2003,wang_threshold_2009,fowler_surface_2012,terhal_quantum_2015,roffe_quantum_2019}. Furthermore, it provides established theoretical foundations for achieving a universal set of logical gates through resource states and patch deformations~\cite{bravyi_universal_2005,horsman_surface_2012,litinski_game_2019,chamberland_universal_2022,gidney_magic_2024}. These appealing features have led to steady progress in experimental demonstrations~\cite{google_quantum_ai_exponential_2021,google_quantum_ai_suppressing_2023,google_quantum_ai_and_collaborators_quantum_2025,eickbusch_demonstrating_2025,he_experimental_2025}, verifying that can be experimentally suppressed via scaling the code distance. To date, efforts to maximize error-suppression efficiency have focused on enhancing decoders, such as by integrating neural networks or utilizing soft information~\cite{bausch_learning_2024,google_quantum_ai_and_collaborators_quantum_2025}, while the Clifford gate sequences used for syndrome extraction have remained predominantly static.

Beyond requiring optimized decoding strategies, realizing long-lived logical qubits in practice requires reducing the architectural demands of building quantum processors via either increasing encoding rates~\cite{bravyi_high-threshold_2024} or reducing coupling map degree. Recent QEC literature has demonstrated how to embed logical qubits under reduced connectivity constraints while preserving full syndrome measurement. These methods generally fall into three categories. First, some approaches utilize heavy-hexagon connectivity with flag qubits to mediate long-range interactions, which usually require deeper syndrome extraction circuits~\cite{chamberland_topological_2020,kim_design_2022,benito_comparative_2025}. Second, stabilizer measurements can be decomposed into non-commuting operators, allowing detectors to be defined statically or time-dynamically without requiring long-range connectivity, though this typically comes at the cost of temporal overhead~\cite{gidney_fault-tolerant_2021,chamberland_topological_2020}. Third, dynamic codes on a hexagonal-lattice can relax structural hardware demands while still extracting the full syndrome in a single round without requiring trade-offs in circuit depth or temporal distance~\cite{mcewen_relaxing_2023,eickbusch_demonstrating_2025}. 

A fundamental limitation of these approaches is their assumption of a defect-free physical architecture, as real processors suffer from defective components, such as non-functional qubits or couplers~\cite{auger_fault-tolerance_2017,mcewen_relaxing_2023,siegel_adaptive_2023,strikis_quantum_2023,benito_comparative_2025}. Fortunately, recent progress in fault-tolerant circuit design has enabled dynamic frameworks, such as LUCI diagrams, to actively avoid syndrome extraction around defective or broken components~\cite{debroy_luci_2025,anker_optimized_2025,wolanski_automated_2026}. While these dynamic codes can still maintain space-time distance despite hardware defects, they accommodate this flexibility by incurring a penalty in timelike distance, specifically requiring multiple measurement rounds to extract the full syndrome~\cite{anker_optimized_2025}. Additionally, experimental demonstration of such frameworks remains unexplored, and their relative performance compared to standard blueprints has not been verified. Lastly, the existing literature focuses on the practical necessity of bypassing defective components, which neglects the more common experimental reality of inhomogeneity. 

In this work, we bridge the aforementioned gaps by experimentally demonstrating dynamic surface codes with an almost halved syndrome density on a superconducting quantum processor. We define syndrome density as the ratio of measured syndrome bits to the total possible per measurement cycle. We achieve targeted suppression of logical bit- or phase-flip errors for the LUCI framework and provide a direct performance comparison against standard code implementations. This establishes the feasibility of QEC architectures on defective hardware, even when operating at a halved temporal distance, and highlights the utility of the framework in actively avoiding highly noisy components.

\begin{figure*}[ht]
    \centering
    \includegraphics[width=0.95\textwidth]{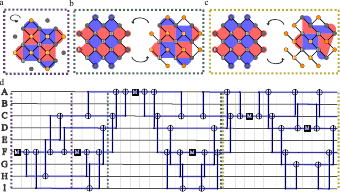}
    \caption{ \textbf{Implementation of surface and LUCI codes and Z detection region.} Distance-3 patches for the standard surface code, the LUCI baseline, and a LUCI variant. Data qubits encoding the logical state are denoted by grey circles, while ancilla qubits measured for syndrome extraction are represented by orange circles. Snapshots show the stabilizer groups of \textbf{a} the standard surface code, alongside the mid-cycle and end-cycle states of \textbf{b} the LUCI framework and \textbf{c} its variant. \textbf{d}. Iterative syndrome extraction rounds for the standard surface and LUCI codes involve gate layers, represented by dotted lines, using the corresponding colors used in \textbf{a}, \textbf{b}, and \textbf{c}. The blue line illustrates how the detection region evolves over consecutive rounds, tracing the operator expansion from $Z_F \rightarrow Z_CZ_DZ_HZ_I \rightarrow Z_AZ_CZ_DZ_F$ to form the mid-cycle state of the LUCI framework. This detection region is contracted onto a single $Z$ measurement, either $Z_A$ in the LUCI baseline or $Z_C$ in the LUCI variant.
    }
    \label{fig1}
\end{figure*}

Here, we consider the LUCI framework, a dynamic code that utilizes a mid-cycle state of the surface code~\cite{debroy_luci_2025}, focusing on its application to rotated surface codes throughout this work. By dynamically altering the measurement scheduling, the LUCI diagrams offer a structural approach to relaxing hardware constraints, but reduce the syndrome density as a consequence. To demonstrate compatibility with contemporary quantum devices, we introduce reset-free circuit implementations of this framework on an IBM superconducting processor (\textit{ibm\_miami})~\cite{geher_reset_2025}. To suppress specific types of logical errors, we scale the QEC patch asymmetrically~\cite{chamberland_universal_2022,higgott_improved_2023}. We establish a distance-3 baseline for both the standard surface code and the LUCI code, and subsequently extend the patches either horizontally ($d_x=5, d_z=3$) or vertically ($d_x=3, d_z=5$) to target logical $X$ or $Z$ error reduction, respectively. For the standard surface code, the syndrome density is $1$ regardless of patch size. In contrast, under the LUCI framework, Equation \eqref{eq:syndrome_density} yields a density of $0.6$ for the distance-3 patch and $0.588$ for the grown patch. The standard surface code achieves suppression ratios ($\Lambda^{5/3}_{\text{Standard}} = \epsilon^r_{d=3}/\epsilon^r_{d=5}$) of $1.58(13)$ and $2.44(7)$ for logical X and Z errors, respectively. In contrast, the LUCI code yields corresponding ratios ($\Lambda^{5/3}_{\text{LUCI}}$) of $1.75(10)$ and $1.93(12)$. Here, $\epsilon^r_3/\epsilon^r_{5}$ denotes the ratio of a specific logical error rate per syndrome measurement round between the $d=3$ and a rectangular patch. Importantly, we focus on a defect-free region for a fair comparison with the standard surface code and explore the impact of nearly halving the temporal syndrome density.

The standard architecture exhibits clear error suppression even when forced to incorporate a highly noisy coupler, with an error rate $\sim 15\%$, in the $X$ basis patch due to the noise inhomogeneity of the processor. In contrast, the dynamic scheduling of the LUCI framework avoids this specific noisy component, yielding superior overall logical error rates while preserving logical boundaries. These findings suggest that standard syndrome extraction circuits may not always be the optimal choice for logical qubit preservation, even in the absence of defects. Instead, it demonstrates that dynamic syndrome extraction can be an alternative approach for maximizing performance on hardware with inhomogeneous noise while maintaining logical boundaries, despite the associated trade-off in temporal syndrome density.

The remainder of the paper is organized as follows. First, we provide brief details on detection region formation within the LUCI framework and discuss syndrome density in section~\ref{syn_extract}. In section~\ref{reset_free}, we discuss how detectors can be defined in the reset-free setup. In section~\ref{hardware}, we outline the hardware implementation and asymmetric patch scaling on the IBM processor. The main results of our work are found in section~\ref{lers}, showing evaluated logical performance for both surface codes and the LUCI framework and highlighting error suppression for targeted logical errors in both approaches. Finally, in section~\ref{discussion} we discuss our findings and promising future directions for practical utility of the LUCI framework.

\section{Flexibility in syndrome extraction}
\label{syn_extract}

Realization of surface codes requires syndrome extraction, which maps commuting $X$- and $Z$-type stabilizers onto single-qubit measurements, projecting the code into an eigenstate of stabilizers over consecutive cycles. By tracking these measurement outcomes across spacetime, detection regions are established,  detecting when errors occur and facilitating active correction~\cite{mcewen_relaxing_2023}. Translating these mappings into Clifford gate sequences has conventionally required devices to realize square-lattice coupling maps to natively support the connectivity for weight-4 stabilizers in the bulk~\cite{arute_quantum_2019,krinner_realizing_2022,marques_logical-qubit_2022,zhao_realization_2022}. However, recent insights have demonstrated the flexibility in syndrome circuit designs employing mid-cycle measurements to establish dynamic codes, an attractive concept across various QEC architectures, including the surface codes~\cite{higgott_handling_2025,debroy_luci_2025, wolanski_automated_2026}. 

The framework for dynamic codes can measure stabilizers around hardware defects to form adaptable detection regions, a task that has been challenging for the standard approach, which embeds a logical qubit using rigid syndrome extraction on a square lattice. Conceptually, this alternative method separates the full stabilizer group into distinct subsets, defining anti-commuting gauge operators when necessary, and employs adaptable operation scheduling to measure the full syndrome group over multiple rounds, an approach known as the LUCI framework~\cite{debroy_luci_2025}. We define each subroutine circuit that measures syndrome bits as a round, which consists of four CNOT gate timesteps, each constituting a gate layer, and a final measurement. A cycle denotes the sequence of rounds required to extract the full syndrome. 

Fig.~\ref{fig1} illustrates how detection regions are reshaped over time in the absence of defects, with Fig.~\ref{fig1} (a) specifically detailing the stabilizer groups for a distance-3 standard surface code, (b) the same distance of the LUCI framework, and (c) the LUCI variant. A cycle in the standard surface code consists of a single round, illustrated by black dotted lines in Fig.~\ref{fig1} (d). In contrast, a LUCI cycle for a patch of the same distance comprises two rounds. The baseline and variant rounds are separated by dotted lines, respectively. Notably, a full LUCI cycle requires ten gate layers in total, double the gate layers of the standard code: eight CNOT gate timesteps and two measurements.

In the standard surface code, whenever ancilla qubits (orange circles) are measured during a syndrome extraction cycle, the adjacent data qubits (grey circles) encoding the logical state are stabilized by the same stabilizer group, as shown in Fig.~\ref{fig1} (a). Conversely, because the gate sequences for the LUCI scheme conclude with CNOT gates while alternating the measurement of stabilizer groups, the framework utilizes a set of stabilizers evolved by these operations. This results in two distinct states, illustrated in Fig.~\ref{fig1} (b) and (c): a mid-cycle state and an end-cycle state. The key insight behind defining the mid-cycle state is that the surface code state can switch to a variant during the implementation of Clifford gates, and vice versa~\cite{mcewen_relaxing_2023}. Hence, alternative surface code circuits can be constructed by measuring mid-cycle stabilizers instead. This mid-cycle state maintains the same code distance as the standard approach, while employing more qubits and over twice the number of stabilizers. Although the cycle is designed to begin and end in this mid-cycle state, the stabilizers temporarily align with those of the standard surface code during every second measurement reached at the end of the cycle. Consequently, this establishes the capability to dynamically transition between the LUCI baseline and the standard approach.

The LUCI baseline employs the same checkerboard pattern as the standard approach for assigning ancilla and data qubits, depicted in Fig.~\ref{fig1} (b). However, there is freedom in choosing which qubits to measure per weight-4 stabilizer in the mid-cycle state, thereby enabling the design of LUCI variants. This flexibility yields $2^{2d(d-1)}$ possible variants for the distance $d$ patch. Fig.~\ref{fig1} (c) illustrates one such variant, constructed by consistently selecting the opposite qubits relative to the baseline.

Introducing additional qubits and applying two layers of CNOT gates transitions the standard surface code into the LUCI mid-cycle state. This allows the circuit to switch between the standard and LUCI syndrome extraction rounds. Fig.~\ref{fig1} (d) isolates a local nine-qubit neighborhood to show how the detection region for the weight-4 $Z$ stabilizer, $Z_CZ_DZ_HZ_I$, changes across spacetime. We omit the contraction phase for the standard surface code to highlight the distinct spacetime volume traced out across the LUCI rounds. As the circuit executes, the equivalent detection region evolves into $Z_AZ_CZ_DZ_F$ in the LUCI framework. In the first round of a cycle, this detection region contracts to a single $Z$ measurement, targeting either $Z_A$ in the baseline or $Z_C$ in the variant. The following CNOT gates expand the region back to $Z_AZ_CZ_DZ_F$. Notably, the first two CNOT layers of the second round temporarily deform the detection region into either $Z_CZ_DZ_HZ_I$ or $Z_AZ_BZ_FZ_G$ to form the mid-cycle state, before it returns to $Z_AZ_CZ_DZ_F$. This sequence continues iteratively across all subsequent cycles. For further details on other stabilizers, we provide the circuit implementing our constructions for dynamically transitioning between the surface code and the LUCI frameworks in \href{https://algassert.com/crumble#circuit=Q(0,1)0;Q(0,2)1;Q(0.5,0.5)2;Q(0.5,1.5)3;Q(0.5,2.5)4;Q(1,0)5;Q(1,1)6;Q(1,2)7;Q(1,3)8;Q(1.5,0.5)9;Q(1.5,1.5)10;Q(1.5,2.5)11;Q(2,0)12;Q(2,1)13;Q(2,2)14;Q(2,3)15;Q(2.5,0.5)16;Q(2.5,1.5)17;Q(2.5,2.5)18;Q(3,1)19;Q(3,2)20;POLYGON(0,0,1,0.25)3_10_11_4;POLYGON(0,0,1,0.25)9_16_17_10;POLYGON(0,0,1,0.25)5_9_2;POLYGON(0,0,1,0.25)11_18_15;POLYGON(1,0,0,0.25)10_17_18_11;POLYGON(1,0,0,0.25)2_9_10_3;POLYGON(1,0,0,0.25)16_19_17;POLYGON(1,0,0,0.25)3_4_1;TICK;R_5_7_13_15_1_6_14_19;MARKZ(0)7;MARKZ(1)6;MARKZ(2)14;MARKZ(3)13;TICK;H_1_6_14_19;TICK;CX_2_5_4_7_10_13_6_3_14_11_19_17;TICK;CX_9_5_11_7_17_13_6_2_14_10_19_16;TICK;CX_3_7_9_13_11_15_1_4_6_10_14_18;TICK;CX_10_7_16_13_18_15_1_3_6_9_14_17;TICK;H_1_6_14_19;TICK;M_8_5_13_7_12_15_0_6_19_14_1_20;TICK;H_0_6_19_14_1_20;TICK;CX_2_5_10_13_4_7_6_3_19_17_14_11;TICK;CX_9_5_6_2_19_16_3_0_17_13_14_10_11_7_8_4;TICK;POLYGON(0,0,1,0.25)6_10_7_3;POLYGON(0,0,1,0.25)13_17_14_10;POLYGON(0,0,1,0.25)12_16_13_9;POLYGON(0,0,1,0.25)5_9_6_2;POLYGON(0,0,1,0.25)7_11_8_4;POLYGON(0,0,1,0.25)14_18_15_11;POLYGON(0,0,1,0.25)5;POLYGON(0,0,1,0.25)12;POLYGON(0,0,1,0.25)8;POLYGON(0,0,1,0.25)15;POLYGON(1,0,0,0.25)10_14_11_7;POLYGON(1,0,0,0.25)3_7_4_1;POLYGON(1,0,0,0.25)17_20_18_14;POLYGON(1,0,0,0.25)16_19_17_13;POLYGON(1,0,0,0.25)9_13_10_6;POLYGON(1,0,0,0.25)2_6_3_0;POLYGON(1,0,0,0.25)0;POLYGON(1,0,0,0.25)1;POLYGON(1,0,0,0.25)19;POLYGON(1,0,0,0.25)20;TICK;TICK;CX_16_12_13_9_10_6_7_3_20_17_4_1_18_14_15_11;TICK;CX_9_12_3_6_11_14_13_10_7_4_20_18;TICK;H_1_13_7_20_0_19;TICK;M_15_12_6_14_8_5_1_13_7_20_0_19;TICK;H_1_13_7_20_0_19;TICK;CX_9_12_3_6_11_14_13_10_7_4_20_18;TICK;CX_16_12_13_9_10_6_7_3_20_17_4_1_18_14_15_11;TICK;TICK;CX_9_5_6_2_19_16_3_0_17_13_14_10_11_7_8_4;TICK;CX_2_5_10_13_4_7_6_3_19_17_14_11;TICK;H_0_6_19_14_1_20;TICK;M_8_5_13_7_12_15_0_6_19_14_1_20;POLYGON(0,0,1,0.25)3_10_11_4;POLYGON(0,0,1,0.25)9_16_17_10;POLYGON(0,0,1,0.25)5_9_2;POLYGON(0,0,1,0.25)11_18_15;POLYGON(1,0,0,0.25)10_17_18_11;POLYGON(1,0,0,0.25)2_9_10_3;POLYGON(1,0,0,0.25)16_19_17;POLYGON(1,0,0,0.25)3_4_1;TICK;H_0_6_19_14_1_20;TICK;CX_2_5_10_13_4_7_6_3_19_17_14_11;TICK;CX_9_5_6_2_19_16_3_0_17_13_14_10_11_7_8_4;POLYGON(0,0,1,0.25)6_10_7_3;POLYGON(0,0,1,0.25)13_17_14_10;POLYGON(0,0,1,0.25)12_16_13_9;POLYGON(0,0,1,0.25)5_9_6_2;POLYGON(0,0,1,0.25)7_11_8_4;POLYGON(0,0,1,0.25)14_18_15_11;POLYGON(0,0,1,0.25)5;POLYGON(0,0,1,0.25)12;POLYGON(0,0,1,0.25)8;POLYGON(0,0,1,0.25)15;POLYGON(1,0,0,0.25)10_14_11_7;POLYGON(1,0,0,0.25)3_7_4_1;POLYGON(1,0,0,0.25)17_20_18_14;POLYGON(1,0,0,0.25)16_19_17_13;POLYGON(1,0,0,0.25)9_13_10_6;POLYGON(1,0,0,0.25)2_6_3_0;POLYGON(1,0,0,0.25)0;POLYGON(1,0,0,0.25)1;POLYGON(1,0,0,0.25)19;POLYGON(1,0,0,0.25)20;TICK;TICK;CX_16_12_13_9_10_6_7_3_20_17_4_1_18_14_15_11;TICK;CX_12_9_6_3_14_11_10_13_4_7_18_20;TICK;H_1_10_4_18_0_19;TICK;M_15_9_3_11_8_5_1_10_4_18_0_19;TICK;H_1_10_4_18_0_19;TICK;CX_12_9_6_3_14_11_10_13_4_7_18_20;TICK;CX_16_12_13_9_10_6_7_3_20_17_4_1_18_14_15_11;TICK;TICK;CX_9_5_6_2_19_16_3_0_17_13_14_10_11_7_8_4;TICK;CX_5_2_13_10_7_4_3_6_17_19_11_14;TICK;H_0_3_17_11_1_20;TICK;M_8_2_10_4_12_15_0_3_17_11_1_20;POLYGON(0,0,1,0.25)9_16_19_13;POLYGON(0,0,1,0.25)6_13_14_7;POLYGON(0,0,1,0.25)5_9;POLYGON(0,0,1,0.25)14_18;POLYGON(1,0,0,0.25)5_9_13_6;POLYGON(1,0,0,0.25)13_19_18_14;POLYGON(1,0,0,0.25)16_19;POLYGON(1,0,0,0.25)6_7;TICK;H_0_3_17_11_1_20;TICK;CX_5_2_13_10_7_4_3_6_17_19_11_14;TICK;CX_9_5_6_2_19_16_3_0_17_13_14_10_11_7_8_4;POLYGON(0,0,1,0.25)6_10_7_3;POLYGON(0,0,1,0.25)13_17_14_10;POLYGON(0,0,1,0.25)12_16_13_9;POLYGON(0,0,1,0.25)5_9_6_2;POLYGON(0,0,1,0.25)7_11_8_4;POLYGON(0,0,1,0.25)14_18_15_11;POLYGON(0,0,1,0.25)5;POLYGON(0,0,1,0.25)12;POLYGON(0,0,1,0.25)8;POLYGON(0,0,1,0.25)15;POLYGON(1,0,0,0.25)10_14_11_7;POLYGON(1,0,0,0.25)3_7_4_1;POLYGON(1,0,0,0.25)17_20_18_14;POLYGON(1,0,0,0.25)16_19_17_13;POLYGON(1,0,0,0.25)9_13_10_6;POLYGON(1,0,0,0.25)2_6_3_0;POLYGON(1,0,0,0.25)0;POLYGON(1,0,0,0.25)1;POLYGON(1,0,0,0.25)19;POLYGON(1,0,0,0.25)20}{Crumble}\unskip, a tool for visualizing circuits and detection regions~\cite{gidney_stim_2021,mcewen_relaxing_2023}.

To quantify the temporal frequency of syndrome extraction, we define syndrome density, $\rho(d_x,d_z)$, as the ratio of measured syndrome bits to the total syndrome bits over rounds per cycle, yielding a value between $0$ and $1$. While the surface code extracts all possible stabilizers within a patch every cycle (or a round), achieving a density of $1$, the LUCI approach requires multiple rounds that measure only a partial set of the stabilizer group, resulting in a density of less than $1$. Specifically, each round of the LUCI framework does not measure exactly half of the stabilizers; rather, it additionally measures a subset of the boundary stabilizers. For the LUCI framework on a rectangular patch, having $d_x$ and $d_z$, regardless of the specific baseline or variant, the syndrome density is given by 
\begin{equation}
\rho(d_x, d_z) = \frac{1}{2} + \frac{d_x + d_z - 2}{2(2d_x d_z + d_x + d_z - 4)},
\label{eq:syndrome_density}
\end{equation}
asymptotically converging to $0.5$ as the patch scales. Consequently, the distance-3 and rectangular patches, either ($d_x=5, d_z=3$) or ($d_x=3, d_z=5$), studied in this work have the syndrome densities of $0.6$ and $0.588$, respectively.

\section{Reset-Free Detectors}
\label{reset_free}
While two-qubit gates and measurements are essential primitives for fault-tolerant circuits, physical reset operations remain an optional architectural feature~\cite{geher_reset_2025}. Syndrome cycle implementations generally fall into three distinct approaches. First, unconditional resets actively return ancillas to the ground state, which makes room to use additional components to remove leakage, enhancing overall circuit efficiency; as a trade-off, this forces inactive qubits to idle during the reset duration~\cite{magnard_fast_2018,zhou_rapid_2021,miao_overcoming_2023}. Second, conditional bit-flip feedback offers the same definition of standard detection regions across spacetime without physical resets, but similarly forces the quantum processor to idle while waiting for classical control logic~\cite{corcoles_exploiting_2021,chen_calibrated_2022,sundaresan_demonstrating_2023}. The third approach omits physical initialization entirely~\cite{marques_logical-qubit_2022,krinner_realizing_2022,miao_overcoming_2023}. While this avoids the qubit idle time that occurs in the other approaches, it requires redefining the standard detection regions, relying on classical post processing. Here, we focus exclusively on this reset-free scenario, detailing how its detectors are constructed within the LUCI framework.

Typically, measurements from two consecutive cycles are sufficient to define detectors in spacetime for the standard surface code with reset gates~\cite{dennis_topological_2002,fowler_surface_2012,gidney_stim_2021}. A detector in the $r$-th cycle, $d_{i,r}$, is defined by the parity of two subsequent measurements of the $i$-th stabilizer: $d_{i,r} = m_{i,r} \oplus m_{i,r+1}$, where $m_{i,r}$ and $m_{i,r+1}$ are outcomes from cycles $r$ and $r+1$. In a reset-free setting, however, this parity information is instead obtained via $d_{i,r} = m_{i,r} \oplus m_{i,r+2}$~\cite{geher_reset_2025}. At the temporal boundaries, the detectors in the first cycle are functionally identical to those in a code using reset gates: $d_{i,1} = m_{i,1}$, provided the stabilizers share the same basis as the initialized logical state. In the final ($n$-th) cycle, the detectors are determined by comparing the last stabilizer outcomes with the data qubit measurements: $d_{i,n} = m_{i,n-1} \oplus m_{i,n} \oplus (\bigoplus_{k \in \text{supp}(i)} m_{q_k})$, where $m_{q_k}$ denotes the measurement of data qubits forming the support of the $i$-th stabilizer.

\begin{figure*}[ht]
    \centering
    \includegraphics[width=0.95\textwidth]{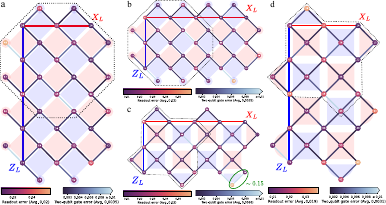}
    \caption{ \textbf{Hardware qubit configuration and code patch mapping.} Circles represent used physical qubits for demonstrations on the IBM quantum processor (\textit{ibm\_miami}), with connecting lines indicating the availability of two-qubit gates ($CZ$). The color bar denotes calibration-derived error rates for measurements and two-qubit gates across the device. Device mappings for the ($d_x=3, d_z=5$) patches are shown for \textbf{a} the LUCI framework and \textbf{d} the surface code in the $X$ basis. Corresponding mappings for the ($d_x=5, d_z=3$) patches are shown for \textbf{c} the surface code and \textbf{b} the LUCI framework in the $Z$ basis. Regions highlighted by black dotted lines give an example of qubits employed for distance-3 experiments in the X (red) and Z (blue) bases. Alternative mapping configurations and qubit layerout number on the device, along with their two-qubit gate and measurement error rates, are detailed in Appendix \ref{app_figures}. Notably, the Z-basis mapping for the standard surface code incorporates a highly noisy coupler, which has an error rate of around $\sim 15\%$ highlighted with the green circle, and is avoided in the corresponding LUCI framework configuration.
    }
    \label{fig2}
\end{figure*}

Every qubit in a LUCI construction actively participates in the contraction and expansion of mid-cycle stabilizers, differing from the standard approach of simply calculating parity outcomes between two static cycles. Hence, rather than using cycles, detectors in the LUCI baseline are defined using measurements from consecutive rounds, such that $d_{i,j} = m_{i,j+1} \oplus \bigoplus_{l \in \text{supp}(i)} m_{q_l,j}$, where $m_{i,j+1}$ is the outcome of the $i$-th stabilizer measurement at rounds $j+1$, and $m_{q_l,j}$ represents the measurements of qubits that support the extended or contracted $i$-th stabilizer at round $j$. 

\begin{figure*}[t]
    \centering
    \includegraphics[width=0.95\textwidth]{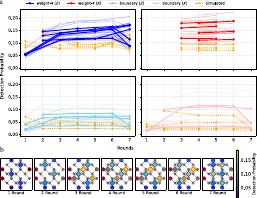}
    \caption{ \textbf{Reset-free detection probability in the distance-3 LUCI baseline.} Illustrating detection event rates and their spatial distribution for a distance-3 LUCI framework. We extract detection event probabilities over 7 rounds using $10^4$ experimental shots in the $Z$ basis. The first two rounds prepare the logical state, while the final round measures the mid-cycle state on the same basis used for the initial preparation. \textbf{a}. Detection probabilities for individual stabilizers over consecutive rounds. Each data point represents the average across three distinct qubit configurations of the distance-3 patch. Weight-4 $X$-type and $Z$-type stabilizers appear as red and blue lines, whereas boundary weight-1 stabilizers appear in pink and skyblue, respectively. Darker lines represent the average across all evaluated qubit configurations, while transparent lines denote the results of individual configurations. \textbf{b}. A spatial heatmap of detection probabilities for all stabilizers across syndrome extraction rounds. The plotted values are averaged across all subgrid configurations. Z stabilizers are indicated by blue-outlined boxes, whereas X stabilizers are denoted by red lines.
    }
    \label{fig3}
\end{figure*}

Remarkably, this framework partially allows the reset-free detector definition to mirror the standard approach with resets. Specifically, parity information can be obtained as $d_{i,r} = m_{i,r} \oplus m_{i,r+1} = m_{i,j} \oplus m_{i,j+2}$. For weight-1 stabilizers, if the stabilizer expands through subsequent gates, its detector is defined identically to those in the bulk; otherwise, it is defined by the parity information of consecutive rounds, $d_{i,j} = m_{i,j} \oplus m_{i,j+1}$. Finally, at the temporal boundaries, the initial detectors are functionally equivalent to an approach using reset gates: $d_{i,1} = m_{i,1}$ and $d_{i,2} = m_{i,2}$, assuming the stabilizers match the basis of the prepared logical state. For the final cycle ($n$) consisting of two rounds, the detectors are resolved as: $d_{i,2n-1} = m_{i,2n-1} \oplus \textstyle \bigoplus_{k \in \text{supp}(i)} m_{q_k}$ and $d_{i,2n} = m_{i,2n} \oplus \textstyle \bigoplus_{k \in \text{supp}(i)} m_{q_k}$. 

\section{LUCI Framework on hardware}
\label{hardware}

We benchmarked the logical memory performance of both the rotated surface code and LUCI frameworks on the \textit{ibm\_miami} processor using data from 15 May 2026. Logical basis states ($\ket{0}_L$ or $\ket{+}_L$) are prepared by initializing all the data qubits in either the $\ket{0}$ ($Z$ basis) or $\ket{+}$ ($X$ basis) state, followed immediately by one syndrome extraction cycle to project the qubits into the codespace. We execute circuits at variable depths up to seven total rounds, including initialization. In the final round, we measure all qubits in the initialization basis to verify the parities of the data qubits and their corresponding stabilizers. This allows us to confirm whether the logical state corrected by the decoder matches the initially prepared state. Notably, for the LUCI framework, this final measurement concludes in the mid-cycle state rather than the end-cycle state. Compared to the standard surface code, terminating in this mid-cycle state requires three additional operational layers: two CNOT gate timesteps and the final measurement. As the measurement duration of approximately $2.4$ $\mu$s is an order of magnitude longer than the two-qubit gates duration of a few hundred nanoseconds, dominating the total operation time, we count this final sequence as one complete round in our results.

We evaluate the logical memory performance of distance-3 patches, as well as asymmetrically grown patches, for both the standard surface code and the LUCI framework, collecting $10^4$ shots per configuration. The LUCI framework configurations are illustrated in Fig.~\ref{fig2} (a) ($d_z = 5, d_x = 3$) and (b) ($d_z = 3, d_x = 5$), while their corresponding configurations for the standard surface code are shown in (d) and (c), respectively. We scale the patches within a defect-free region of the processor, utilizing overlapping physical qubit subsets across both frameworks. Specifically, we implement three distinct distance-3 codes embedded within the rectangular patches, along the data qubit footprint of the fully grown patch. One of the distance-3 patches is shown with the black dotted lines in Fig.~\ref{fig2}. Their qubit layout numbers on hardware and other configurations for the distance-3 are provided in Appendix \ref{app_figures}. 

In the quantum memory experiments in the $Z$ basis, we select qubit configurations with comparable noise levels for both frameworks. The error rates for two-qubit gates and measurements are on the order of $10^{-3}$ and $10^{-2}$, respectively, as summarized in Fig.~\ref{fig2}. Contrastingly, because standard syndrome extraction requires four neighboring qubits, the $X$-basis mapping for the standard surface code is forced to use a highly noisy coupler to match the mid-cycle state boundaries used in the LUCI framework. This coupler, highlighted by a green circle in Fig.~\ref{fig2}(c), has an error rate of $\sim 15\%$, which the LUCI framework avoids. 

These configurations employ up to 29 physical qubits and 44 couplers for the standard surface code, compared to 35 physical qubits and 50 couplers for the LUCI diagrams. In these experiments, the duration of a single syndrome extraction cycle is approximately $3.5\ \mu$s for the standard code and $7\ \mu$s for the LUCI framework. Finally, to mitigate the impact of idling errors on overall logical performance, we implement XY4 dynamical decoupling~\cite{rahman_learning_2024}. Following an As-Late-As-Possible (ALAP) gate scheduling, these decoupling sequences are applied to any qubits left idling, whether between active gate layers or during measurements. 

\begin{figure*}[ht]
    \centering
    \includegraphics[width=0.95\textwidth]{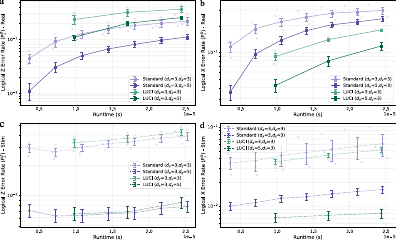}
    \caption{ \textbf{Logical error rates over runtime.} Total logical error rates are plotted as a function of runtime for both the standard surface code and the LUCI baseline, evaluated using the PyMatching decoder. Experimental results, derived from $10^4$ shots without post-selection, are shown for the (a) $X$ basis and (b) $Z$ basis. Darker markers represent the average across all evaluated qubit configurations, while transparent markers denote the results of individual configurations. The corresponding circuit-level noise simulations, parameterized by experimental calibration data using Stim ($10^6$ samples), are displayed for the (c) $X$ basis and (d) $Z$ basis. Uncertainties for the distance-3 data points are based on the standard deviation across individual cases, while those for the rectangular patch are derived from a binomial confidence interval.
    }
    \label{fig4}
\end{figure*}

The detection probability is a good indicator to check how frequently errors occur spatially and temporally during the cycles~\cite{google_quantum_ai_exponential_2021}. In Fig.~\ref{fig3} (a), we show the average detection probabilities for the three distance-3 LUCI baselines in the $Z$ basis, implementing three cycles that include logical state preparation, with the final logical measurement in the mid-cycle state. The same plots for other settings are provided in Appendix~\ref {app_figures}. Each round measures a subgroup of the full syndrome, where two consecutive rounds are required to extract the full syndrome, and contains both weight-4 stabilizers in the bulk and weight-1 stabilizers at the boundaries. 

The bulk weight-4 $X$ and $Z$ stabilizers exhibit average detection event probabilities of approximately $0.149$ and $0.124$, respectively, while the boundary stabilizers have lower probabilities of $0.057$ and $0.055$. Scaling the patch to a larger distance ($d_X=3, d_Z=5$) increases these probabilities to $0.167$ and $0.130$ for the bulk stabilizers, and to $0.066$ and $0.058$ at the boundaries. In the LUCI framework, all detection probabilities increase as the patch is scaled in the $Z$ basis; conversely, in the alternative basis, introducing more qubits decreases the average detection probability.

\section{LUCI vs Surface Code}
\label{lers}
We conduct repeated fault-tolerant rounds for both the surface code ($R\in\{1,2,3,4,5,6,7\}$) and LUCI codes ($R\in\{3,5,7\}$), reaching a total circuit duration of up to $\sim25$ $\mu$s. To target the suppression of specific logical $X$ or $Z$ errors, we test distance-3 codes alongside asymmetric rectangular variants scaled either horizontally ($d_x=5, d_z=3$) or vertically ($d_x=3, d_z=5$), extracting the logical error rate per round. As the LUCI framework alternates gate sequences across the two rounds of a cycle, we report its performance as a per-round average. In this work, we assume that each round contributes equally to the overall logical error. We decode using correlated Minimum Weight Perfect Matching (MWPM) via PyMatching~\cite{fowler_optimal_2013,higgott_sparse_2025}, weighting edges from calibration data for single-, two-qubit gates and measurement. While the MWPM-based decoder is a sub-optimal choice for LUCI codes due to the presence of hyperedges, it achieves competitive performance relative to Belief Propagation with Ordered Statistics Decoding (BP-OSD) results (see Appendix \ref{bposd})~\cite{roffe_decoding_2020}.

We plot the extracted logical $X$ and $Z$ error rates from the processor ($ibm\_miami$), shown in Fig.~\ref{fig4} (a) and (b), alongside Stim simulations under the circuit-level noise~\cite{gidney_stim_2021}, depicted in Fig.~\ref{fig4} (c) and (d). We extract the average logical error rate per round, $\epsilon_r$, by fitting the data to the equation $p_L = \frac{1}{2} (1 - (1 - 2\epsilon_r)^{n})$, where $n$ denotes the number of rounds~\cite{google_quantum_ai_and_collaborators_quantum_2025}. To evaluate performance, we average the error rates across three distinct distance-3 subgrids for the X and Z bases independently. For the distance-5 configurations, we evaluate a single geometry for each target basis: $(d_X=3, d_Z=5)$ and $(d_X=5, d_Z=3)$. The resulting logical error rates for both approaches are summarized in Table \ref{tab:ler}, with uncertainties evaluated from the fit covariance for the separate averaged logical $X$- and $Z$-error fits~\cite{takou_logical_2026}.

\begin{table}[htbp]
    \centering
    \caption{Fitted logical error rates per round for the standard surface code and the LUCI framework. For distance-3 implementations, performance is evaluated by averaging across three distinct subgrids for the $X$ and $Z$ bases independently. Conversely, the distance-5 configurations utilize a single patch tailored to each target logical error suppression: $(d_X=3, d_Z=5)$ for the logical $Z$ error and $(d_X=5, d_Z=3)$ for the logical $X$ error.}
    \label{tab:ler}
    \begin{tabular}{@{}llcc@{}}
        \toprule
        \textbf{Framework} & \textbf{Basis} & $\boldsymbol{\epsilon^r_{d=3}}$ & $\boldsymbol{\epsilon^r_{d=5}}$ \\ 
        \midrule
        \multirow{2}{*}{Standard} & X & $4.25(11)\%$ & $1.74(3)\%$ \\
                                  & Z & $7.54(58)\%$ & $4.76(12)\%$ \\ 
        \addlinespace
        \multirow{2}{*}{LUCI}     & X & $9.02(30)\%$ & $4.67(24)\%$ \\
                                  & Z & $3.11(3)\%$ & $1.77(10)\%$ \\ 
        \bottomrule
    \end{tabular}
\end{table}

Overall, both the standard surface code and the LUCI baseline demonstrate clear error suppression when comparing distance-3 patches with the grown patch in the target basis. Specifically, the error suppression ratio ($\Lambda^{5/3}$) of the logical phase errors is $2.44(7)$ for the standard surface code and $1.93(12)$ for the LUCI baseline. In contrast, the error suppression ratio of the logical bit-flip errors is $1.58(13)$ for the standard surface code and $1.75(10)$ for LUCI. The uncertainties in the error suppression ratios are calculated using standard error propagation from the per-round logical error rates shown in Table \ref{tab:ler}. These results show an asymmetry in logical performance that matches the trend observed in the simulator results. While the standard surface code demonstrates a higher error suppression ratio for logical $Z$ errors, the LUCI framework excels at suppressing logical $X$ errors by leveraging spatial flexibility to bypass highly noisy hardware couplers. Remarkably, this flexibility enables the distance-3 LUCI baseline to achieve a lower logical $X$ error rate than the grown patch of the standard surface code.

We also evaluate a variant of the LUCI baseline that configures measurements to always select qubits with lower measurement errors based on calibration data. This approach does not always yield better logical performance. Altering the measurement configuration can increase the detector volume, making the extracted syndromes less informative and degrading logical performance~\cite{anker_optimized_2025}. Despite this penalty, the variant should still yield logical error rates comparable to the baseline. However, we do not observe the expected error suppression from the experimental demonstrations. We discuss this discrepancy in the discussion and provide the corresponding plots. Supplementary figures detail the total logical error rates evaluated using the BP-OSD decoder in Appendix \ref{bposd} as well as the detection event probabilities in Appendix \ref{app_figures}.

\section{DISCUSSION AND OUTLOOK}
\label{discussion}

In this work, we evaluated the performance of the surface code and LUCI framework stabilizer measurement circuits on the IBM quantum processor (\textit{ibm\_miami}). To evaluate their logical performance, we compared the error suppression ratios achieved by asymmetrically scaling the code distance from $d=3$ to $d=5$ for suppressing targeted Pauli error rates. Implementing the LUCI approach doubles the duration of a single fault-tolerant cycle, from $3.5$ $\mu$s to $7$ $\mu$s, effectively almost halving the temporal syndrome density. Surprisingly, despite this doubled temporal overhead, we still observe clear error suppression in both X and Z bases. In the absence of a highly noisy coupler in the Z basis, the standard surface code outperforms the LUCI baseline. However, when a patch uses a highly noisy component in the X basis, the standard surface code's logical performance degrades significantly, even if some degree of error suppression is maintained. In contrast, the LUCI framework is flexible enough to avoid these noisy couplers; we successfully improve logical performance using the dynamic syndrome measurements while preserving the original logical boundaries via deploying the LUCI codes. These results demonstrate that the LUCI architecture is a promising tool not only for dealing with defects, but also for mitigating localized high-error components by sacrificing the temporal distance.

Although our demonstrations establish the viability of the LUCI framework using an MWPM-based decoder, this choice leaves room for optimization. The dynamic scheduling and mid-cycle states inherent to the LUCI framework introduce hyperedges in the detection graph, making PyMatching not an optimal choice~\cite{debroy_luci_2025}. Future implementations should evaluate and explore advanced decoding algorithms, such as the Relay BP decoder~\cite{muller_improved_2025,maurer_real-time_2025} or neural network approaches~\cite{gicev_scalable_2023,bausch_learning_2024,senior_scalable_2026}. We also highlight that ensemble decoders exhibit strong performance on LUCI circuits in simulation, providing a valuable approach for evaluating their performance in experimental demonstrations~\cite{jones_improved_2024,shutty_efficient_2024}. Beyond these algorithmic improvements, integrating soft information from the analog readout of physical qubits~\cite{bausch_learning_2024} and the characterization of noise in syndrome cycles provides a clear path to achieving higher decoding accuracy~\cite{flammia_averaged_2022,gicev_quantum_2024,hockings_scalable_2025,takou_logical_2026}. Instead of relying solely on binary outcomes, continuous IQ data can be incorporated directly into the decoder or used in post-selection to identify leaked states~\cite{lee_scalable_2026}. We believe that transitioning to such tailored decoders will further enhance the error suppression ratios of dynamic surface codes and maximize their utility against inhomogeneous hardware noise.

Additionally, the LUCI framework requires measuring stabilizers over multiple rounds to obtain the full syndrome, which reduces the temporal distance~\cite{debroy_luci_2025}. Hence, it is intriguing to explore the resulting discrepancy between the spatial and temporal code distances; stability experiments are needed to determine how the temporal distance changes, designing such an experiment for the LUCI framework remains a highly promising avenue for future research~\cite{gidney_stability_2022}.

Furthermore, we investigated a variant of the LUCI framework that consistently selects physical qubits with lower measurement error rates compared to the baseline. Although such an approach does not always yield better results, as it could increase the detection region volume and make the syndrome less informative, we observed that detection probabilities increase significantly across multiple layers regardless of basis and qubit configuration. Hence, we cannot observe error suppression in a LUCI variant setting; its logical error rates are provided in Appendix \ref{bposd}. The primary difference between the LUCI baseline and its variant lies solely in the qubit measurement configuration. Specifically, while the baseline selects measurement qubits in a checkerboard pattern, ensuring that no measured qubits are adjacent, the variant measures a few neighboring qubits simultaneously. Previous literature has reported several instances where qubit measurements induce crosstalk in neighboring qubits~\cite{chen_calibrated_2022,hothem_measuring_2025,gicev_crosstalk_2026}. Therefore, we attribute the performance degradation observed when shifting the measurement configuration primarily to this measurement-induced crosstalk from adjacent qubits. We leave a detailed characterization of this noise for future work. However, modeling this effect will be essential for effectively utilizing the LUCI architecture to address defective processor components.

Finally, we highlight that the LUCI framework offers significant flexibility in allocating physical qubits on a square lattice. The framework allows dynamic reconfiguration to actively select higher-performing local qubits for quantum memory. In the LUCI framework, a logical patch is not restricted to a rigid rectangular shape. It can adopt an alternative physical boundary provided the patch remains topologically equivalent to the target code. This flexibility deforms the shape of local stabilizers but strictly preserves the logical observables. Moving forward, it will be enlightening to evaluate the tradeoffs between the improved logical error rates yielded by these local variants and the complexity of performing adaptable lattice surgery on customized boundaries for logical gate implementation.

The standard surface codes are rigid; they are forced to use the physical qubits they are mapped to, regardless of local error rates. In contrast, the LUCI framework uses dynamic circuits and can actively route around high-noise components. We compared both approaches using numerical simulations and experimental data from the IBM quantum processor (\textit{ibm\_miami}). The results show that the flexibility of the LUCI diagram yields better logical performance than the static standard approach in the presence of highly noisy components in both simulation and experimental demonstrations, while preserving logical boundaries. However, this dynamic framework comes with a caveat: altering the measurement configuration introduces a more complex noise source, such as measurement-induced crosstalk. Making future LUCI variants practical will require characterizing and mitigating this specific noise.

\section*{ACKNOWLEDGMENT}
YK acknowledges the support of the CSIRO Research Training Program Scholarship and the University of Melbourne Research Training Scholarship. The University of Melbourne supported the research through the establishment of the IBM Quantum Network Hub.
\\
\section*{Data availability}
The data that support the findings of this study are included within the article. Further details are available from the corresponding author upon reasonable request.

\section*{Author Contributions}
YK developed and simulated codes under the supervision of MU and MS. YK carried out all experiments and plotted figures with input from MU and SG. YK wrote the manuscript and the appendix with input from all authors.


\twocolumngrid

\beginsupplement

\clearpage
\onecolumngrid
\begin{center}
	\textbf{\large Supplementary information for\\ 
``LUCI on IBM Hardware: Error Suppression with Almost Half Syndrome Density"}
\end{center}

\begin{figure*}[h]
    \centering
    
    \subfloat[Logical $Z$ errors\label{fig:logical_z}]{
        \includegraphics[width=0.48\textwidth]{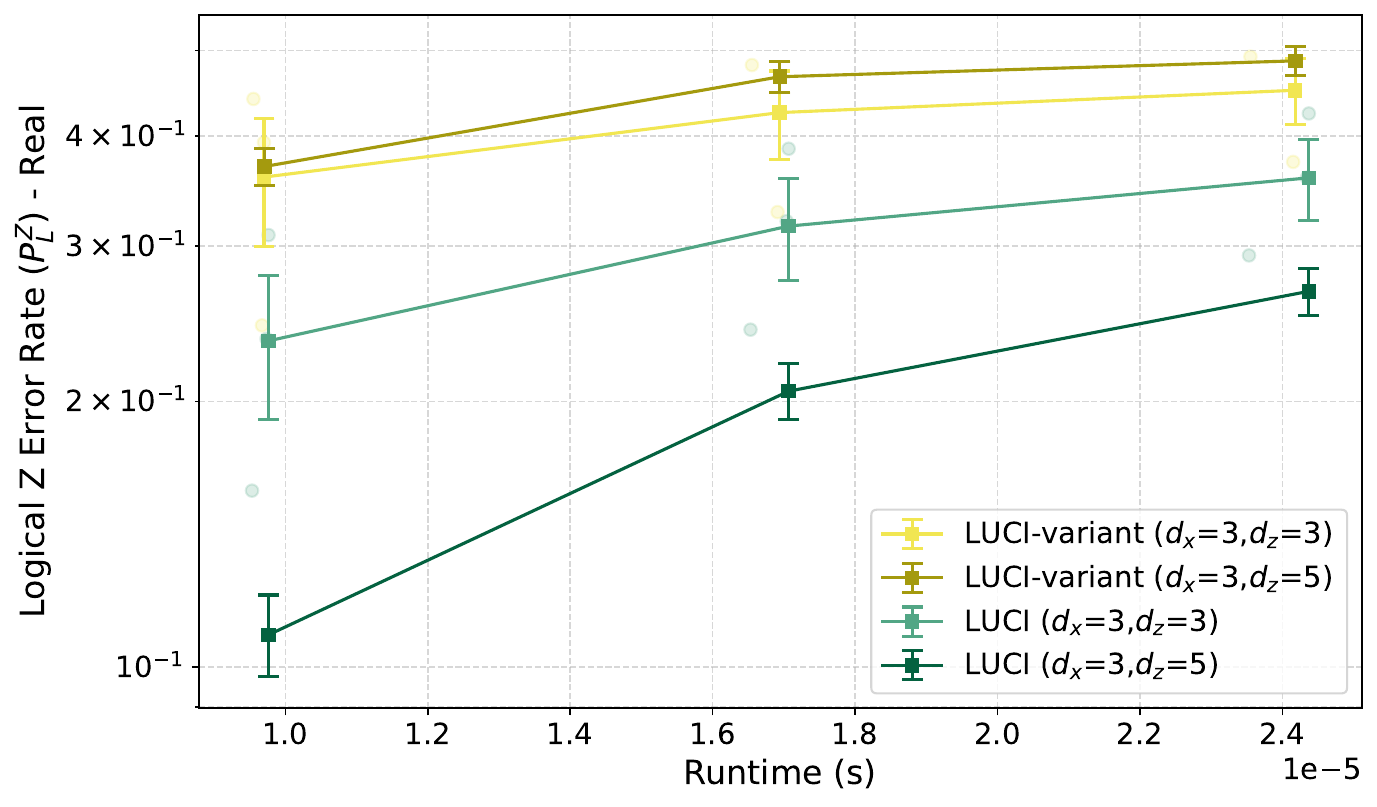}
    }
    \hfill
    \subfloat[Logical $X$ errors\label{fig:logical_x}]{
        \includegraphics[width=0.48\textwidth]{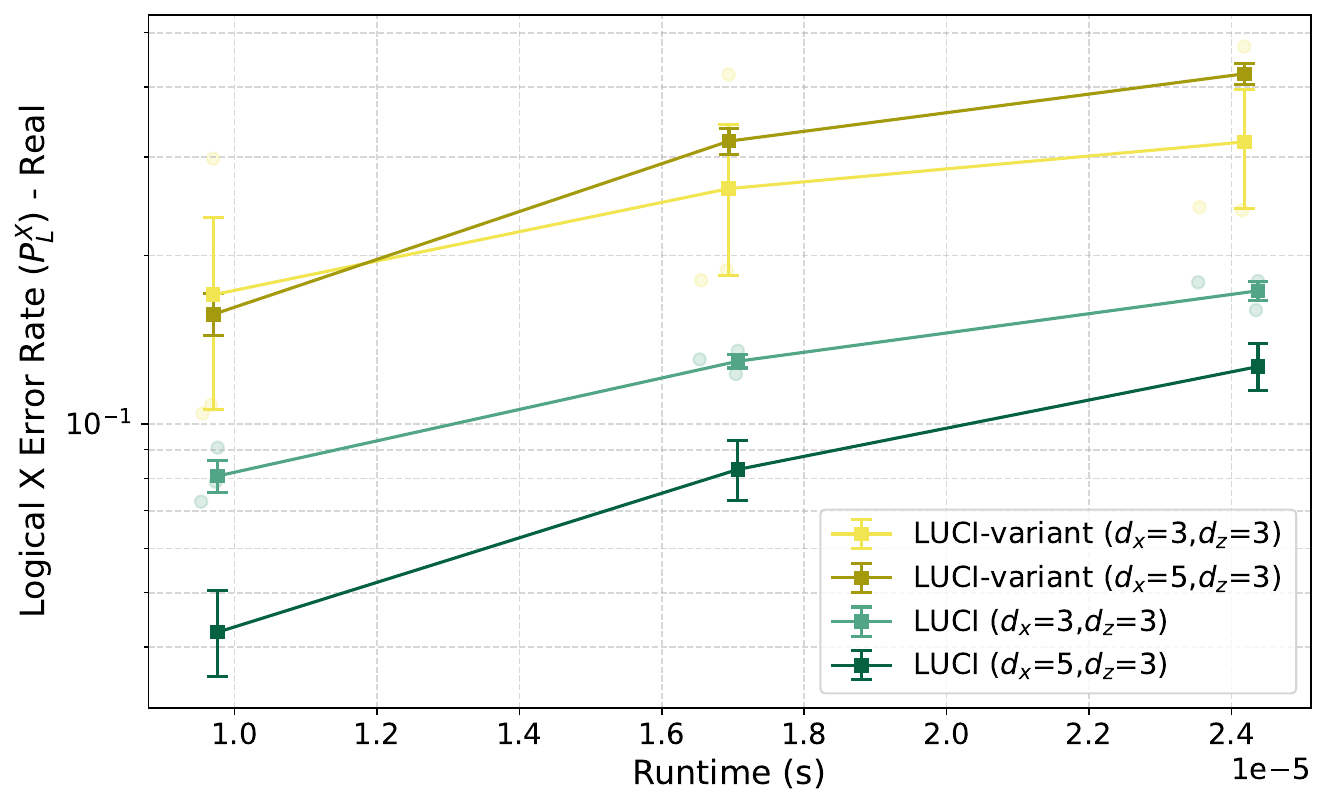}
    }
    
    \caption{We plot the logical error rate as a function of the runtime for both the LUCI baseline and its variant. We decode all cases using the BP-OSD decoder with $10^4$ iterations and OSD order 20. Experimental results, derived from $10^4$ shots without post-selection, are shown for the (a) $X$ basis and (b) $Z$ basis. Darker markers represent the average across all evaluated qubit configurations, while transparent markers denote the results of individual configurations. The uncertainties of the distance-3 codes are calculated via the standard error of the sample mean. Those of the rectangular patch are obtained from the underlying binomial distribution.}
    \label{fig:LER_bposd}
\end{figure*}

\section{Logical Performance of LUCI Diagrams with BP-OSD}
\label{bposd}
To decode the LUCI framework, which inherently involves dynamic measurement scheduling, a decoder algorithm should handle its hyperedges, specifically error mechanisms that create more than two detection events in spacetime. The MWPM-based decoder, implemented in this work via PyMatching, is fundamentally restricted to a graph with edges connecting two nodes, including boundary errors. Therefore, applying this approach to dynamic surface codes requires approximating or decomposing these hyperedges while accounting for correlated errors, which leads to suboptimal performance. In contrast, Belief Propagation with Ordered Statistics Decoding (BP-OSD) operates directly on the Tanner graph representation of the code's parity-check matrix. By iterative message passing, probabilistic soft information between nodes, BP-OSD natively processes hyperedges without requiring structural decomposition, thereby using the spatial and temporal correlation of the errors.

However, the BP-OSD decoder incurs intense computational cost, which currently makes it challenging to use for real-time implementation. In Fig.~\ref{fig:LER_bposd}, we illustrate logical error rates when the BP-OSD decoder, with a maximum of $10^4$ iterations and an OSD order of 20, does not always yield better logical performance than the correlated Pymatching decoder, while it has a significant latency bottleneck. In our evaluations, decoding a single sample with BP-OSD required between $2\times10^{-2}$ s for the distance-3 LUCI baseline code and up to around $9\times10^{-2}$ s for the rectangular patch, whereas PyMatching executed the same task in approximately $2.5 \times 10^{-5}$ s. Interestingly, we couldn't observe a significant improvement in logical error rates regardless of measurement configuration on the LUCI frameworks. This performance gap in time of roughly three orders of magnitude while maintaining comparable logical performance illustrates why an MWPM-based decoder remains the practical choice for the demonstrations. Future work will need to explore timely, accelerated decoders, such as FPGA-based implementations or parallel window-based decoders, to achieve real-time decoding for dynamic circuits, while simultaneously accounting for more complex error mechanisms to improve logical performance.

\section{Further Figures}
\label{app_figures}

\begin{figure*}[h]
    \centering
    
    \subfloat[First distance-3 qubit configuration.\label{fig:std_d3_z_1}]{
        \includegraphics[width=0.3\textwidth]{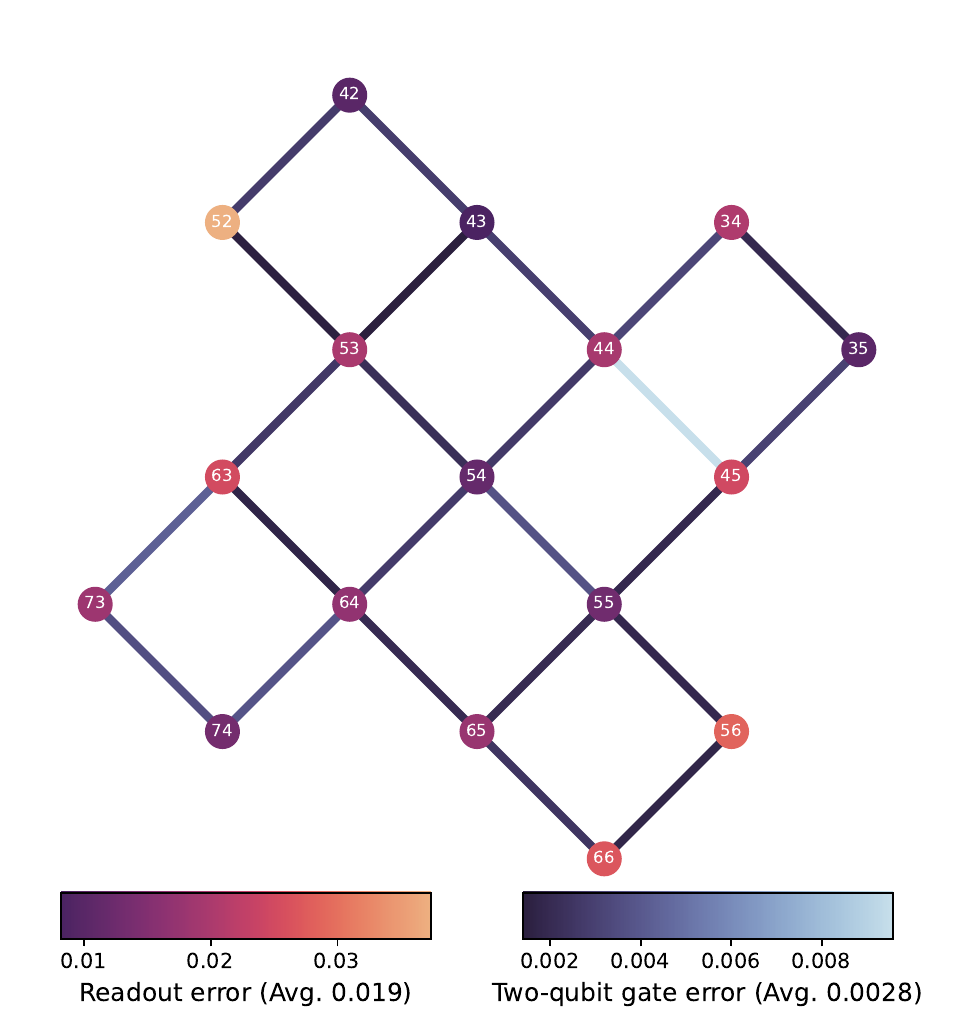}
    }
    \hfill
    \subfloat[Second distance-3 qubit configuration.\label{fig:std_d3_z_2}]{
        \includegraphics[width=0.3\textwidth]{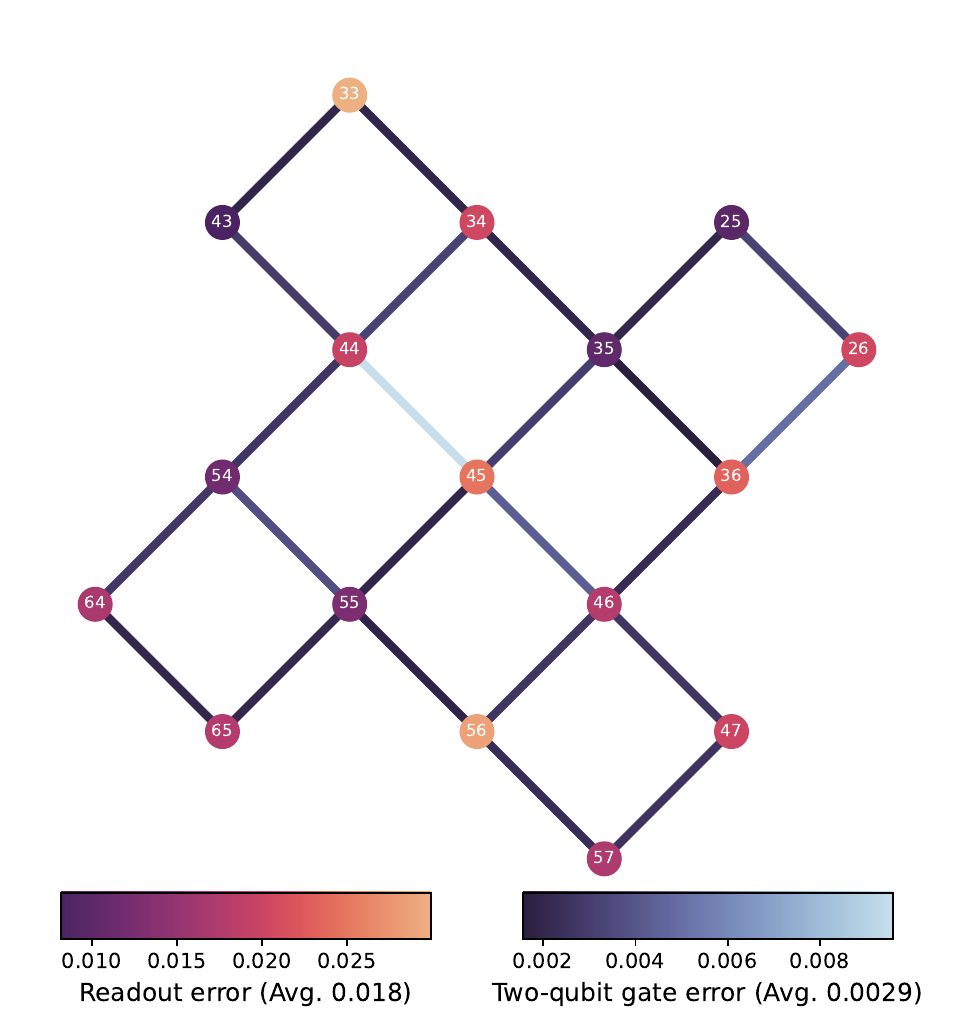}
    }
    \hfill
    \subfloat[Third distance-3 qubit configuration.\label{fig:std_d3_z_3}]{
        \includegraphics[width=0.3\textwidth]{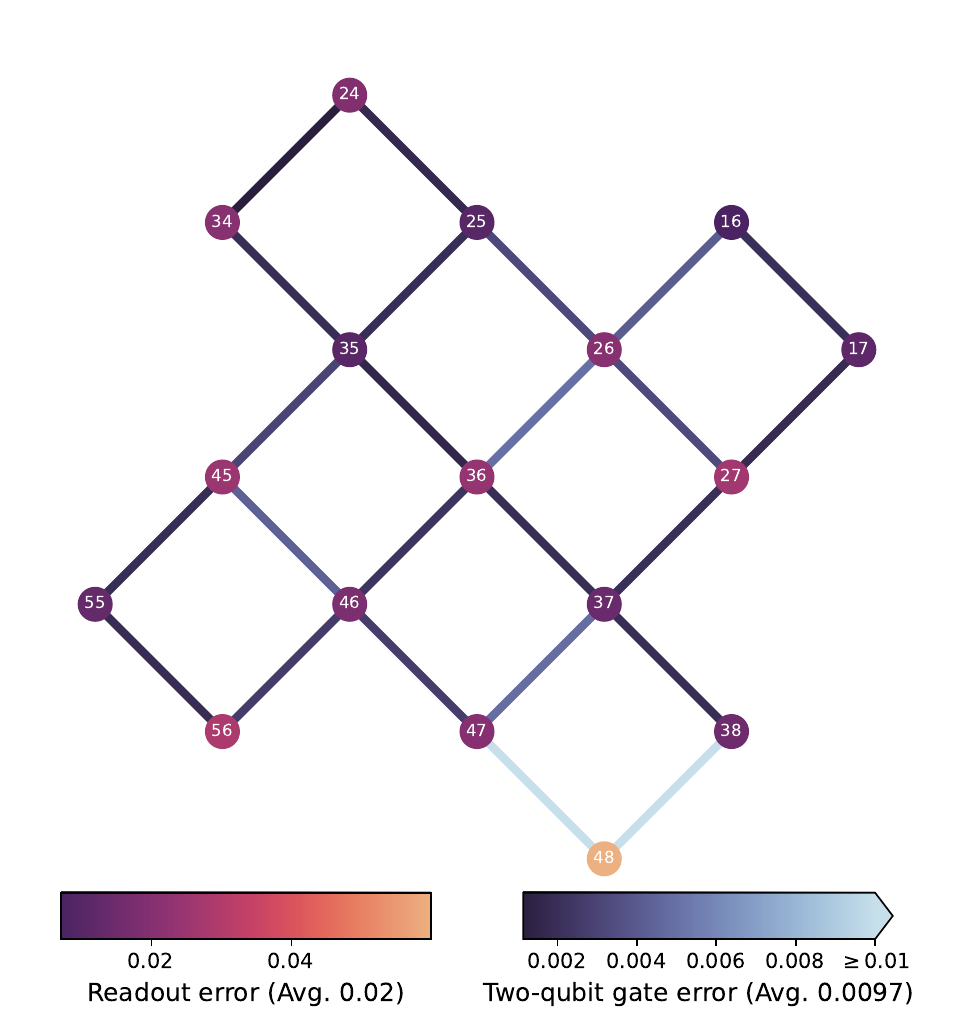}
    }
    
    \vspace{1em}
    
    \subfloat[Detection probabilities.\label{fig:std_d3_z_prob}]{
        \includegraphics[width=\textwidth]{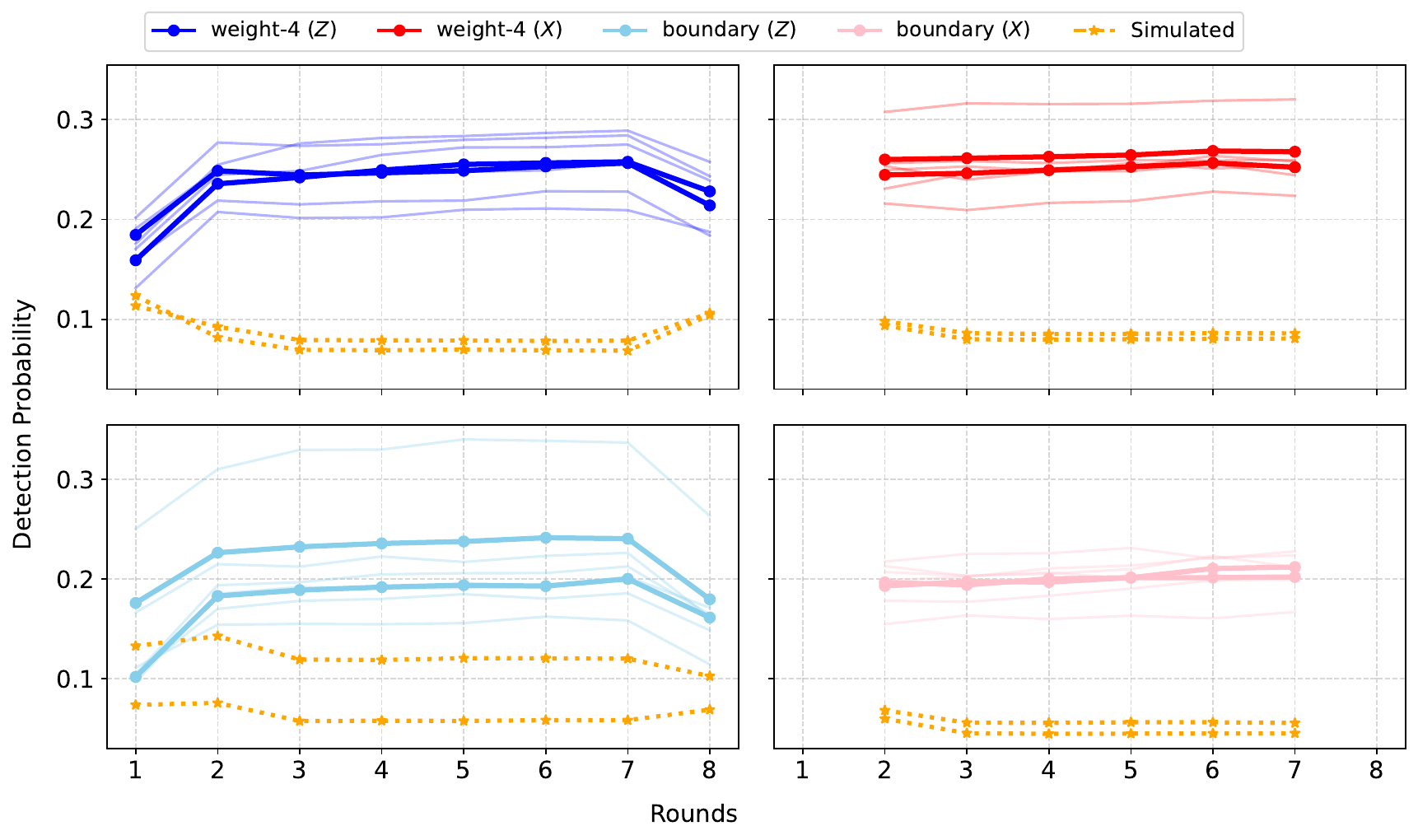}
    }

    \vspace{1em}

    \subfloat[Heatmap of detection probabilities over rounds.\label{fig:std_d3_z_heat}]{
        \includegraphics[width=\textwidth]{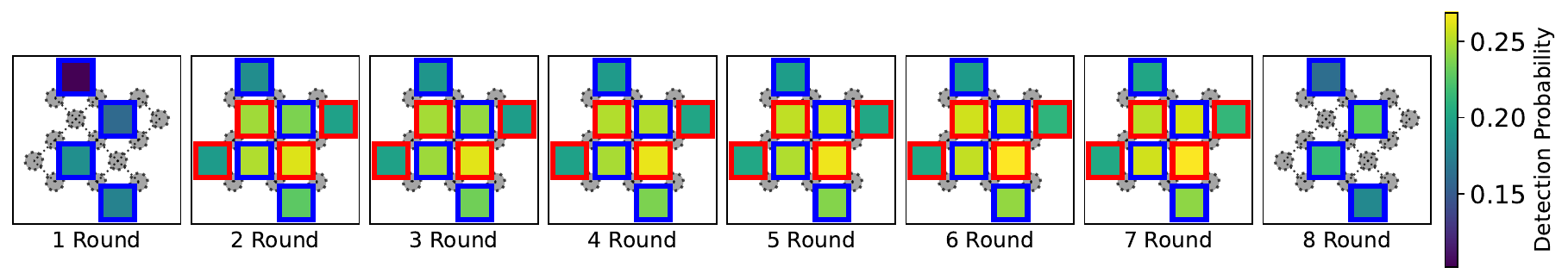}
    }
    
    \caption{Qubit configurations and reset-free detection probabilities for distance-3 surface codes prepared and measured in the Z basis. Results were obtained using $10^4$ samples per patch on the \textit{ibm\_miami} device.}
    \label{fig:surface_code_d3_Z_analysis}
\end{figure*}

\begin{figure*}[h]
    \centering
    
    \subfloat[First distance-3 qubit configuration.\label{fig:std_d3_x_1}]{
        \includegraphics[width=0.3\textwidth]{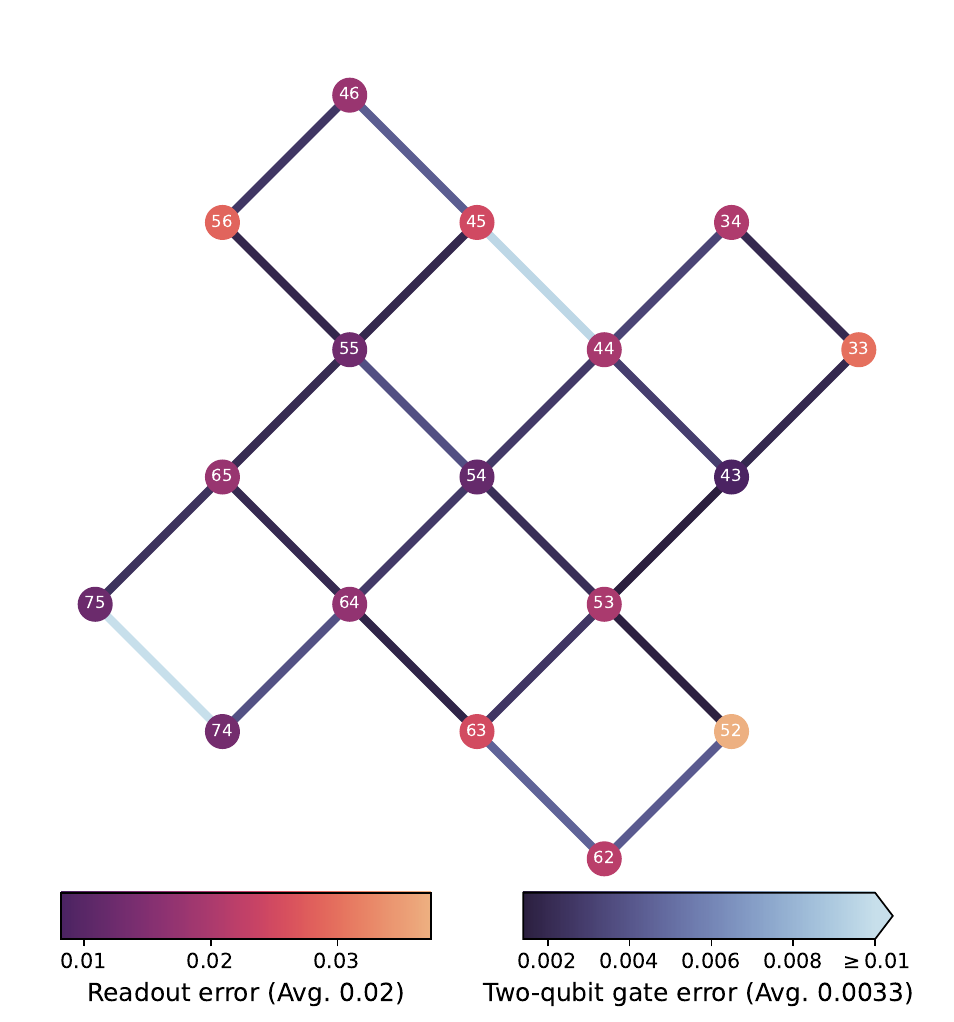}
    }
    \hfill
    \subfloat[Second distance-3 qubit configuration.\label{fig:std_d3_x_2}]{
        \includegraphics[width=0.3\textwidth]{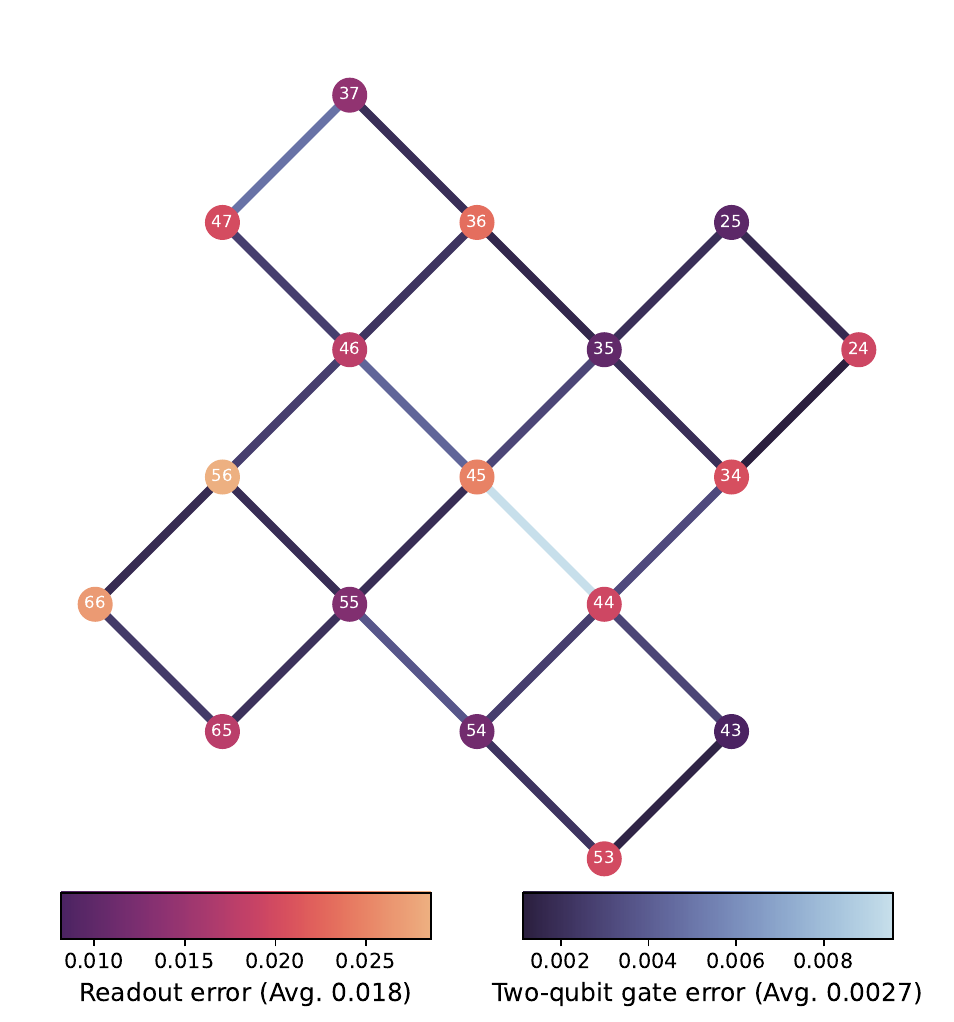}
    }
    \hfill
    \subfloat[Third distance-3 qubit configuration.\label{fig:std_d3_x_3}]{
        \includegraphics[width=0.3\textwidth]{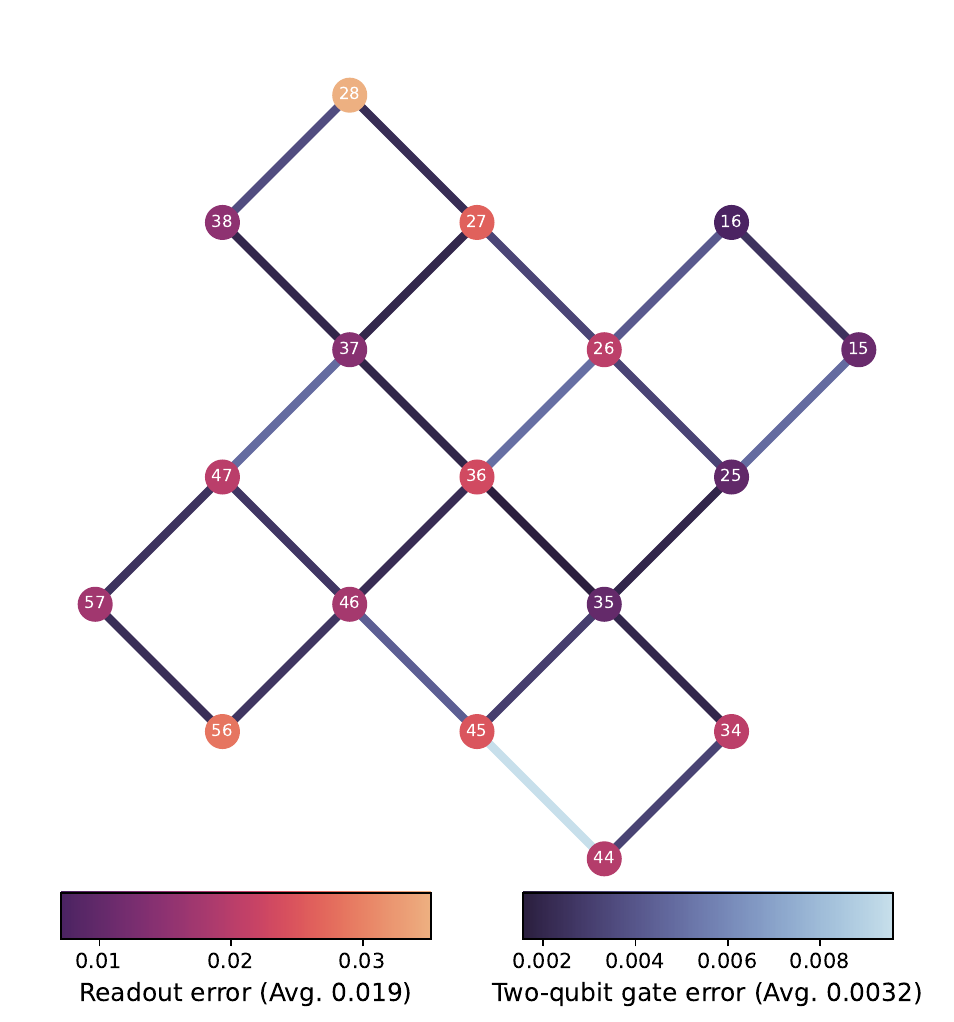}
    }
    
    \vspace{1em}
    
    \subfloat[Detection probabilities.\label{fig:std_d3_x_prob}]{
        \includegraphics[width=\textwidth]{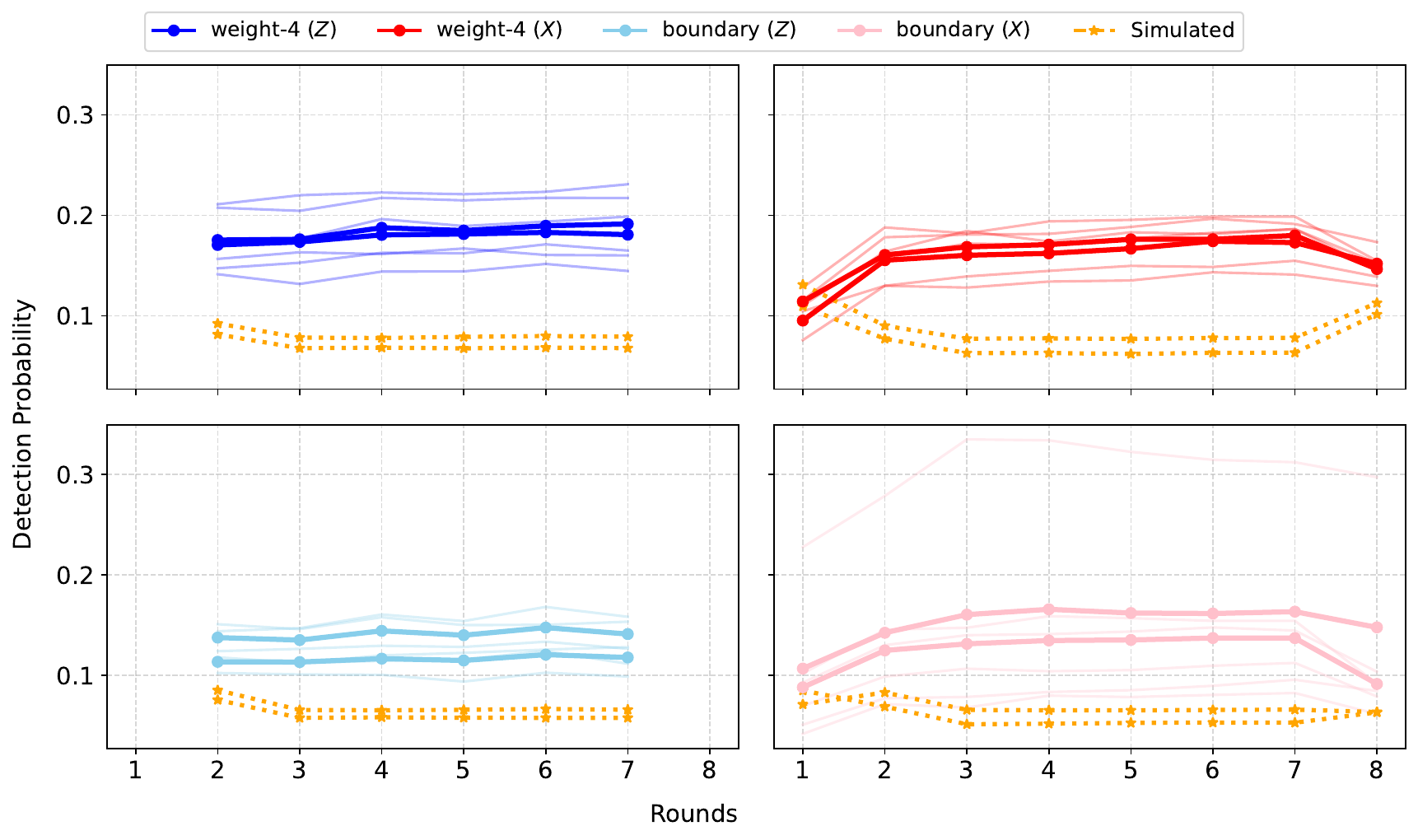}
    }

    \vspace{1em}

    \subfloat[Heatmap of detection probabilities over rounds.\label{fig:std_d3_x_heat}]{
        \includegraphics[width=\textwidth]{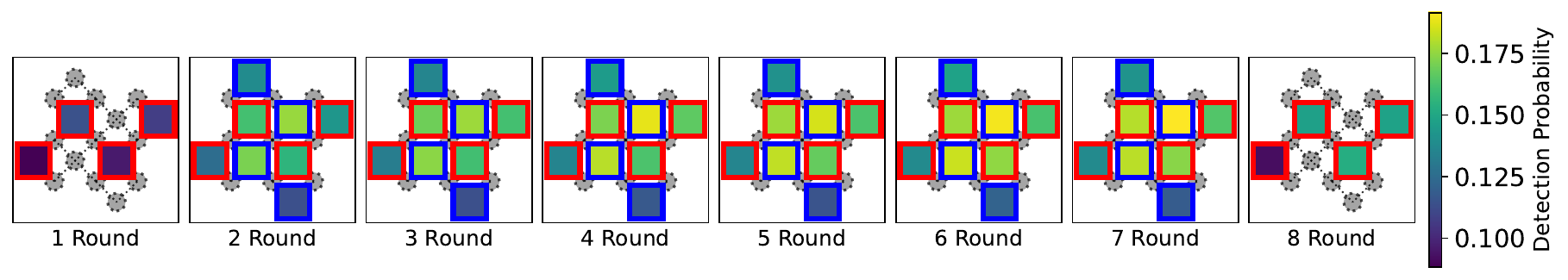}
    }
    
    \caption{Qubit configurations and reset-free detection probabilities for distance-3 surface codes prepared and measured in the X basis. Results were obtained using $10^4$ samples per patch on the \textit{ibm\_miami} device.}
    \label{fig:surface_code_d3_X_analysis}
\end{figure*}

\begin{figure*}[h]
    \centering
    
    \subfloat[$d_X=5$ and $d_Z=3$ qubit configuration.\label{fig:std_d53_z_config}]{
        \includegraphics[width=0.6\textwidth]{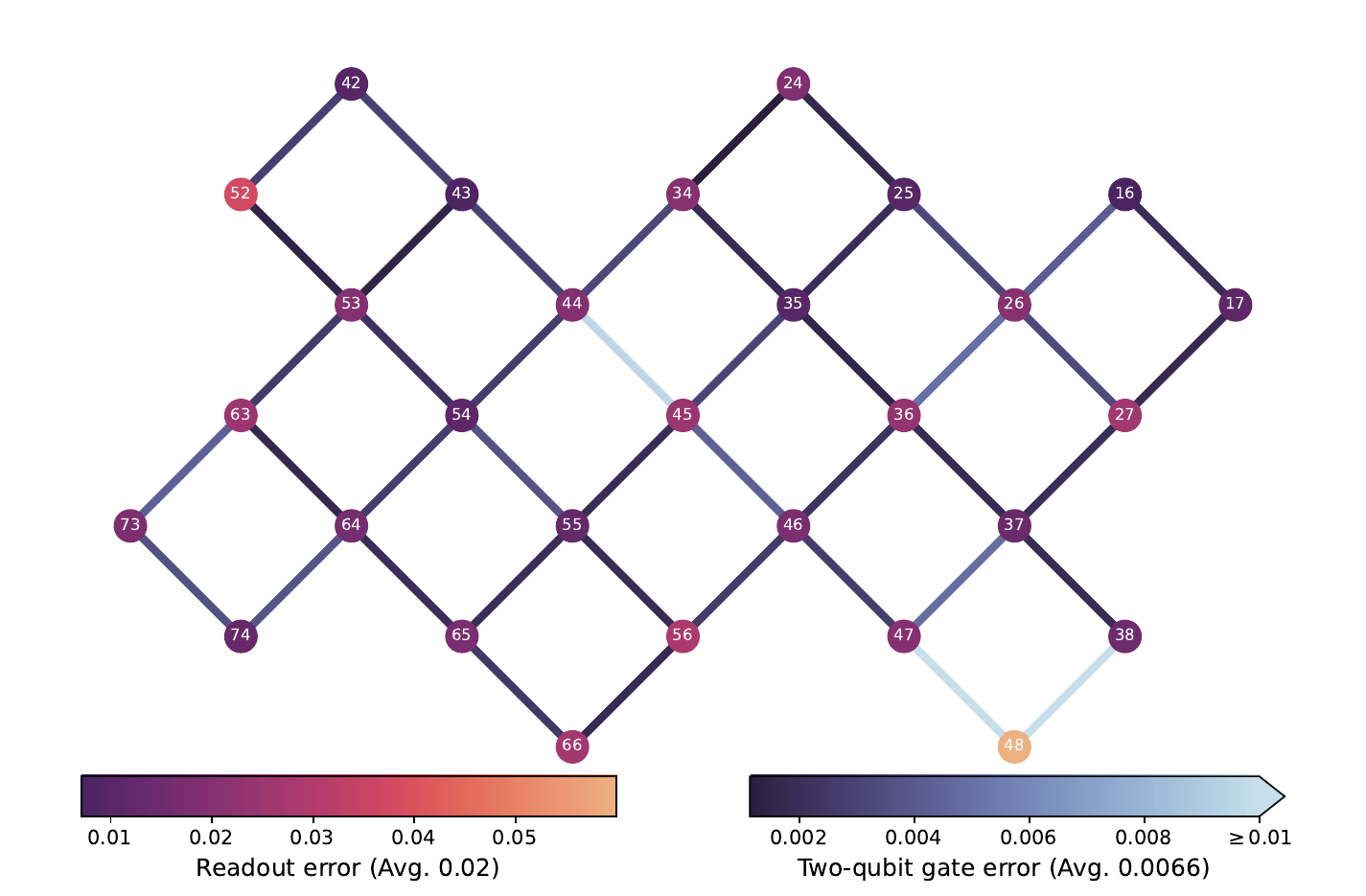}
    }
    
    \vspace{1em}
    
    \subfloat[Detection probabilities.\label{fig:std_d53_z_prob}]{
        \includegraphics[width=0.9\textwidth]{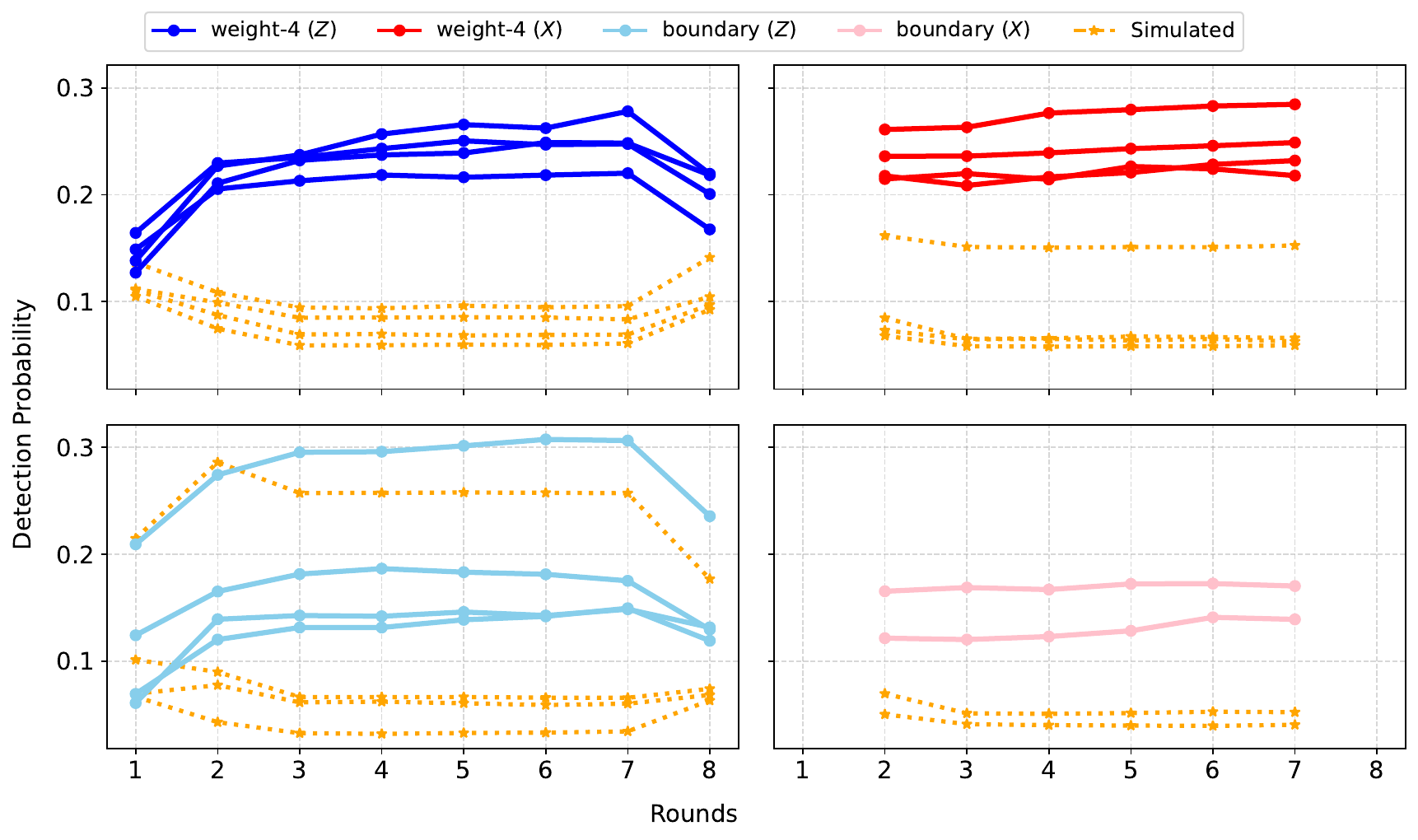}
    }
    
    \vspace{1em}
    
    \subfloat[Heatmap of detection probabilities over rounds.\label{fig:std_d53_z_heat}]{
        \includegraphics[width=\textwidth]{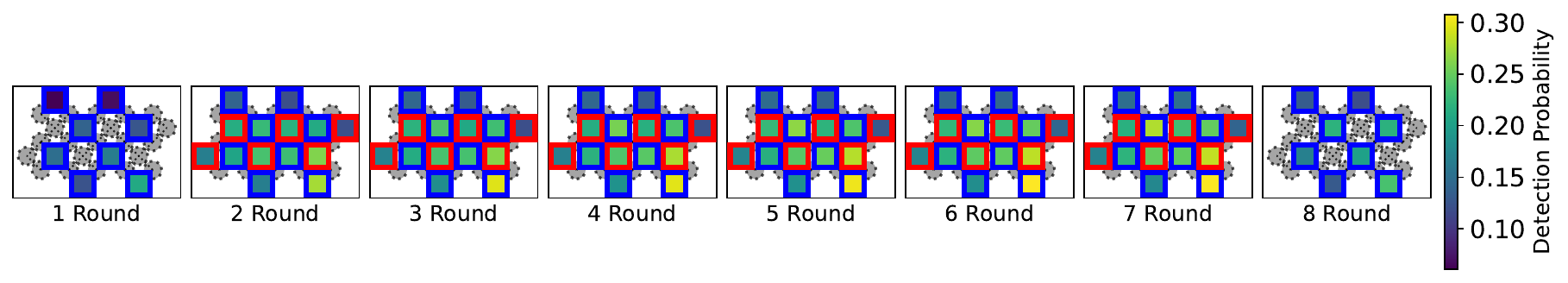}
    }
    
    \caption{Qubit configurations and reset-free detection probabilities for $d_X=5$ and $d_Z=3$ surface code patch prepared and measured in the Z basis. Results were obtained using $10^4$ samples per patch on the \textit{ibm\_miami} device.}
    \label{fig:surface_code_d53_Z_analysis}
\end{figure*}

\begin{figure*}[h]
    \centering
    
    \subfloat[$d_X=3$ and $d_Z=5$ qubit configuration.\label{fig:std_d35_x_config}]{
        \includegraphics[width=0.33\textwidth]{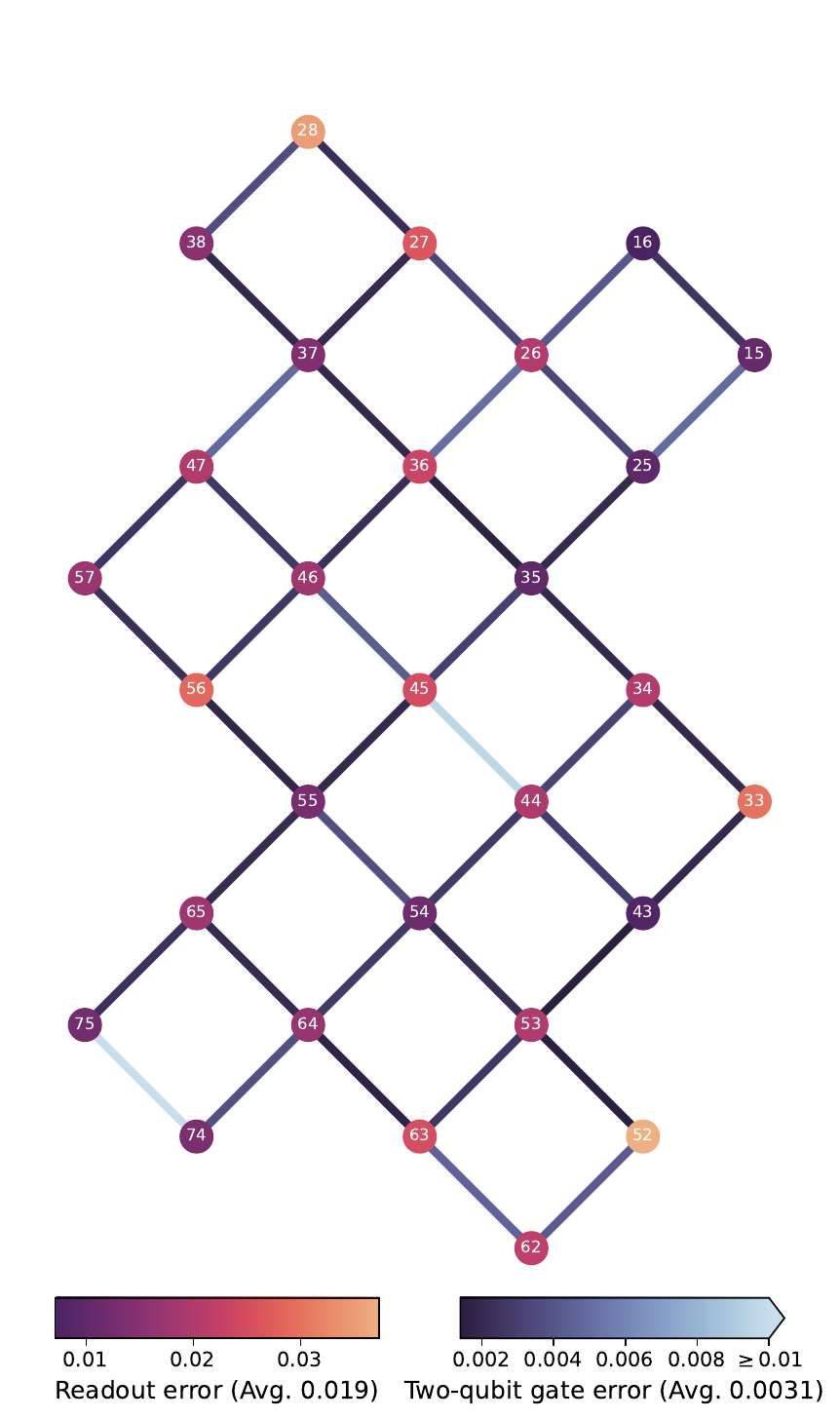}
    }
    \hfill
    \subfloat[Detection probabilities.\label{fig:std_d35_x_prob}]{
        \includegraphics[width=0.62\textwidth]{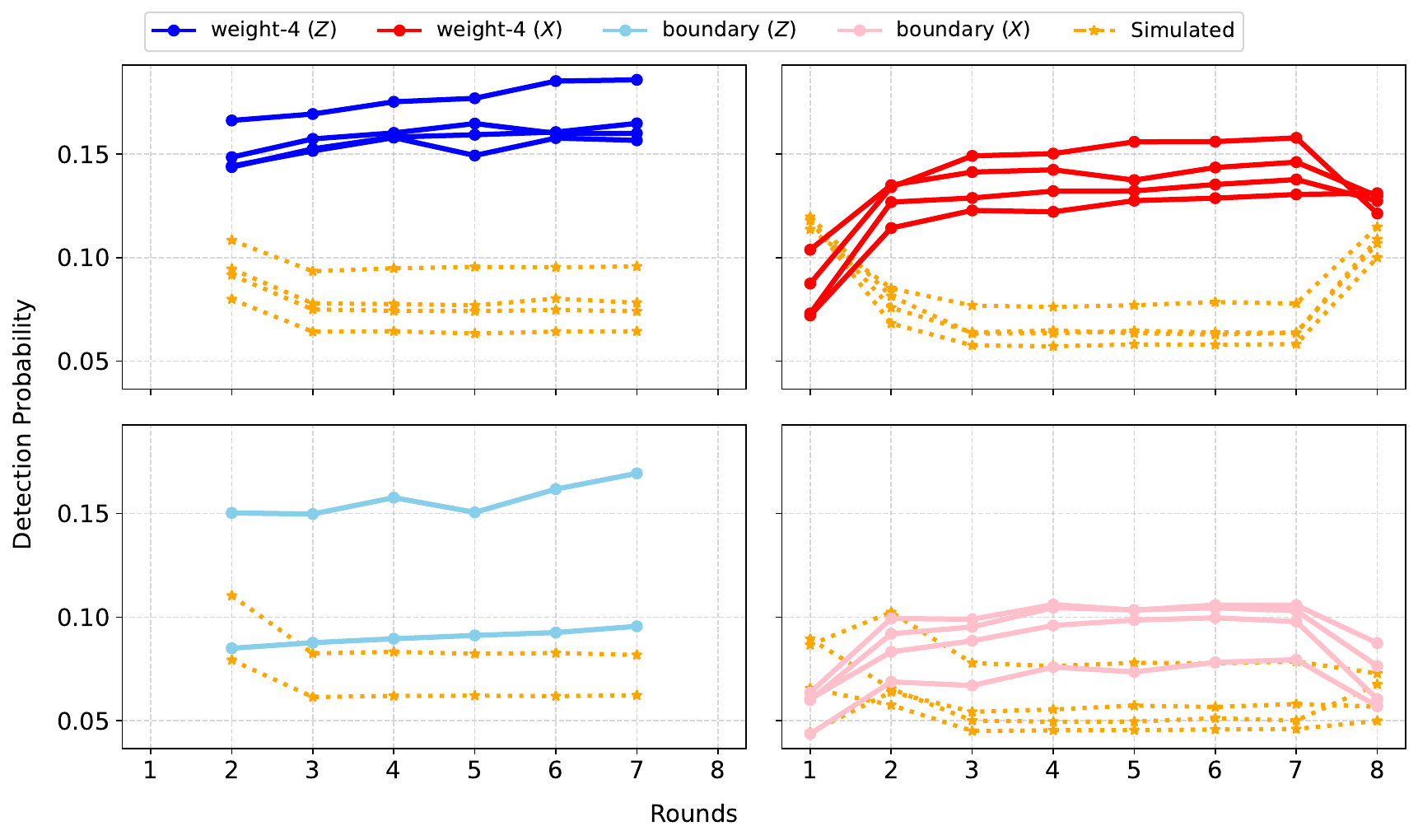}
    }
    
    \vspace{2em}
    
    \subfloat[Heatmap of detection probabilities over rounds.\label{fig:std_d35_x_heat}]{
        \includegraphics[width=\textwidth]{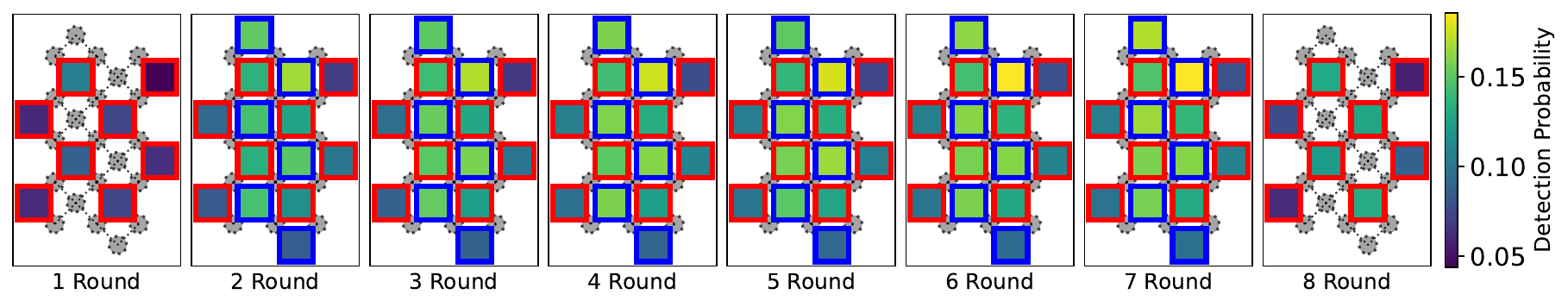}
    }
    
    \vspace{1em}
    
    \caption{Qubit configurations and reset-free detection probabilities for $d_X=3$ and $d_Z=5$ surface code patch prepared and measured in the X basis. Results were obtained using $10^4$ samples per patch on the \textit{ibm\_miami} device.}
    \label{fig:surface_code_d35_X_analysis}
\end{figure*}

\begin{figure*}[h]
    \centering
    
    \subfloat[First distance-3 qubit configuration.\label{fig:acid_d3_z_1}]{
        \includegraphics[width=0.3\textwidth]{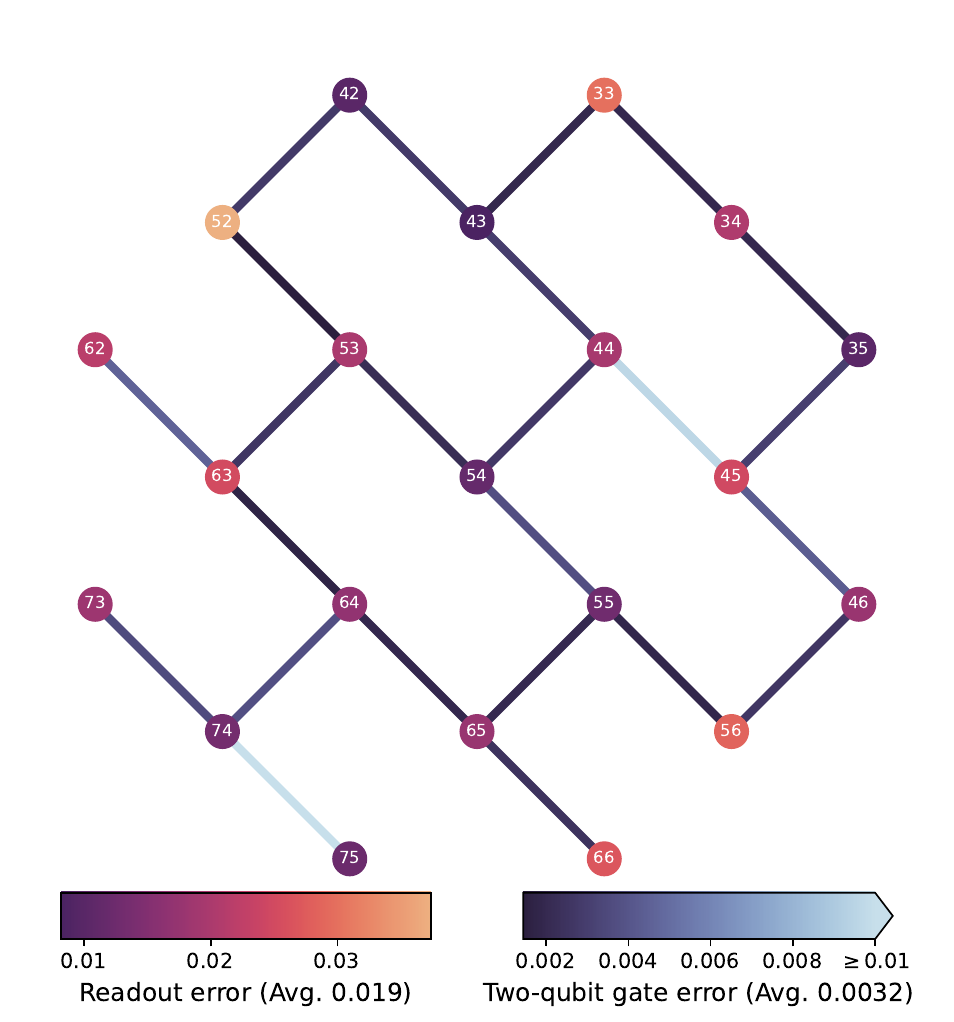}
    }
    \hfill
    \subfloat[Second distance-3 qubit configuration.\label{fig:acid_d3_z_2}]{
        \includegraphics[width=0.3\textwidth]{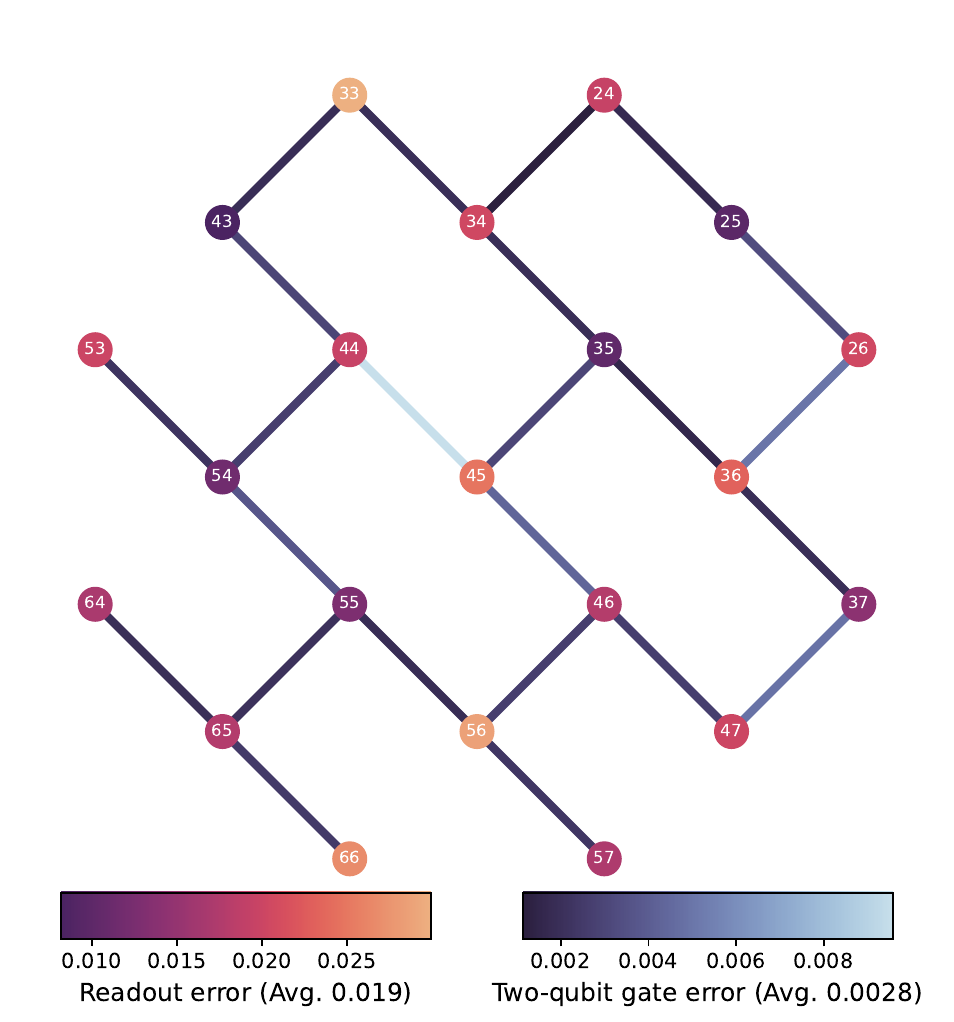}
    }
    \hfill
    \subfloat[Third distance-3 qubit configuration.\label{fig:acid_d3_z_3}]{
        \includegraphics[width=0.3\textwidth]{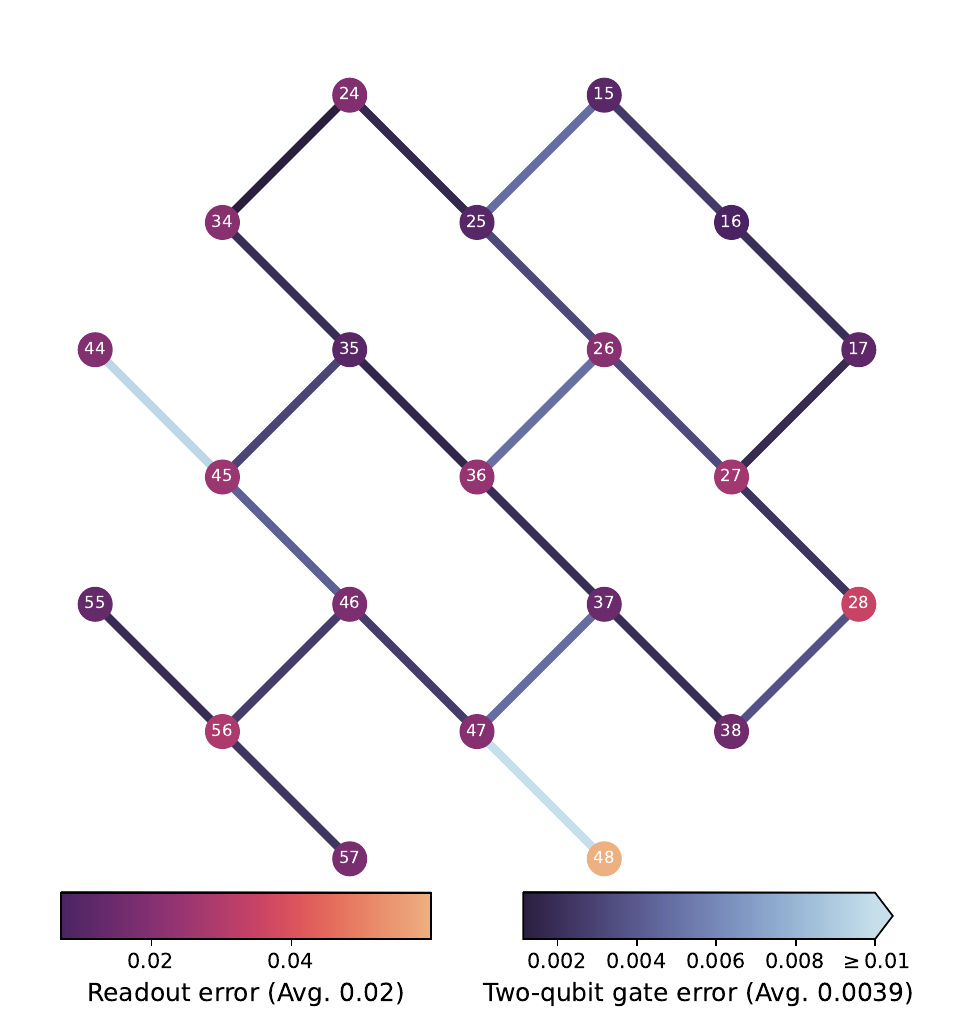}
    }
    
    \vspace{1em}
    
    \subfloat[Detection probabilities.\label{fig:acid_d3_z_prob}]{
        \includegraphics[width=\textwidth]{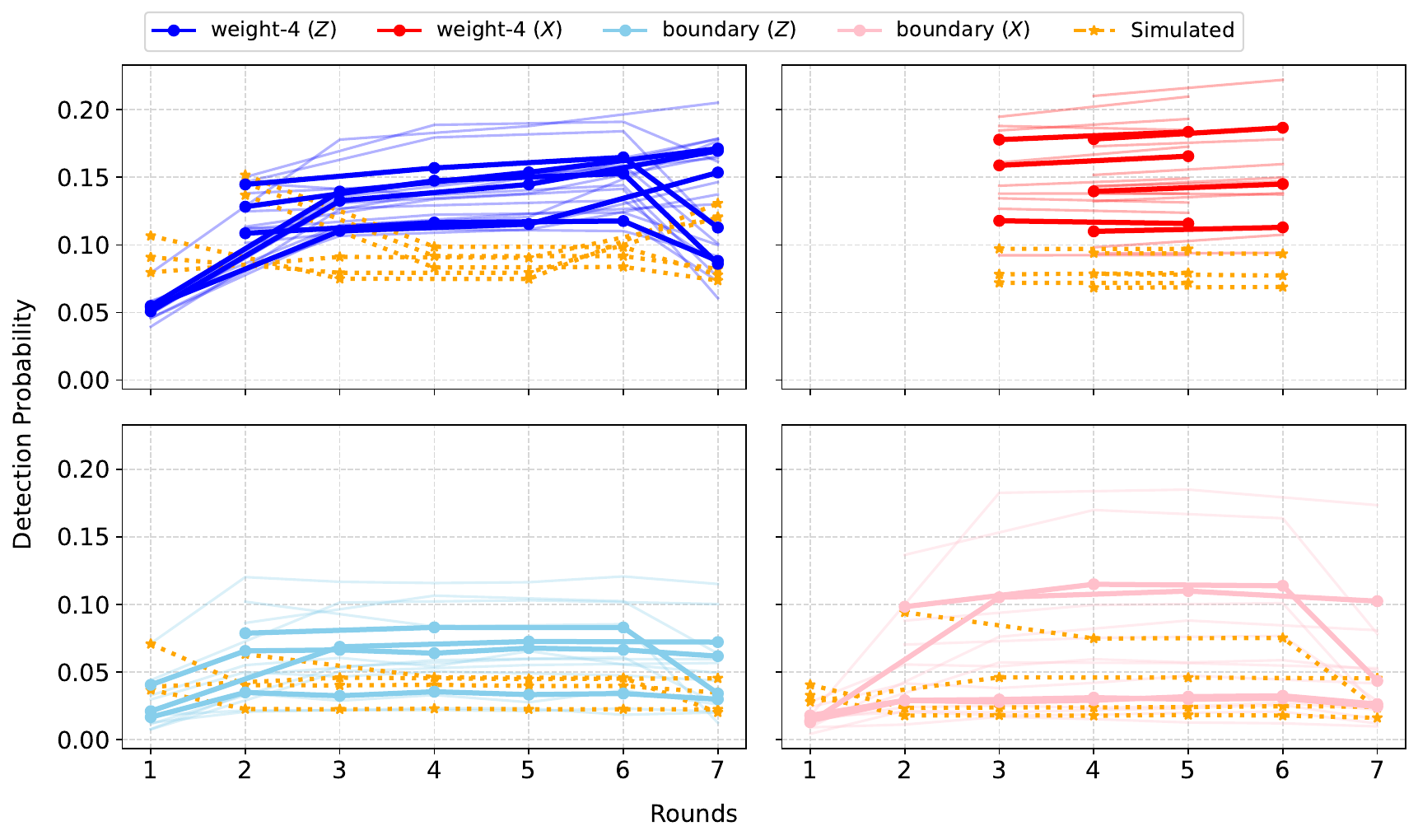}
    }
    
    \vspace{1em}
    
    \subfloat[Heatmap of detection probabilities over rounds.\label{fig:acid_d3_z_heat}]{
        \includegraphics[width=\textwidth]{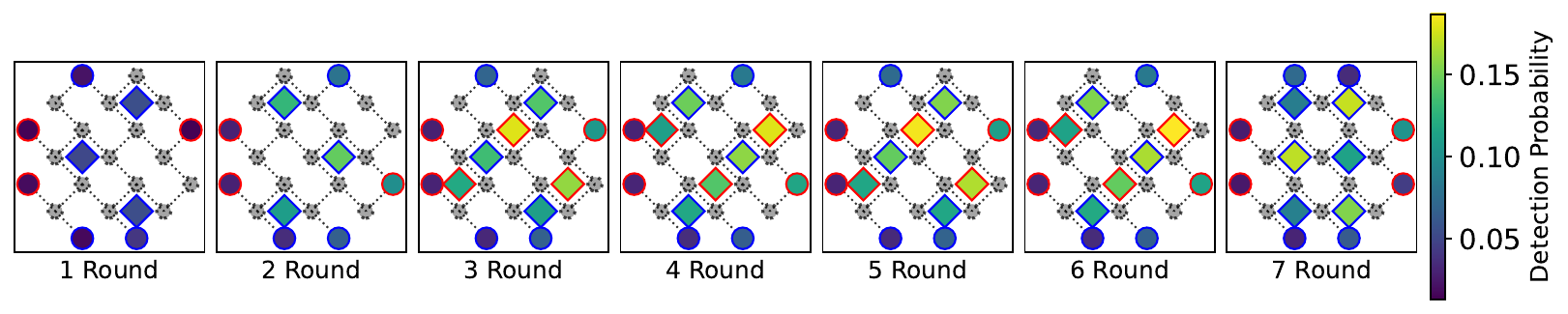}
    }
    
    \caption{Qubit configurations and reset-free detection probabilities for distance-3 LUCI baseline codes prepared and measured in the Z basis. Results were obtained using $10^4$ samples per patch on the \textit{ibm\_miami} device.}
    \label{fig:LUCI_d3_Z_analysis}
\end{figure*}

\begin{figure*}[h]
    \centering
    
    \subfloat[First distance-3 qubit configuration.\label{fig:acid_d3_x_1}]{
        \includegraphics[width=0.3\textwidth]{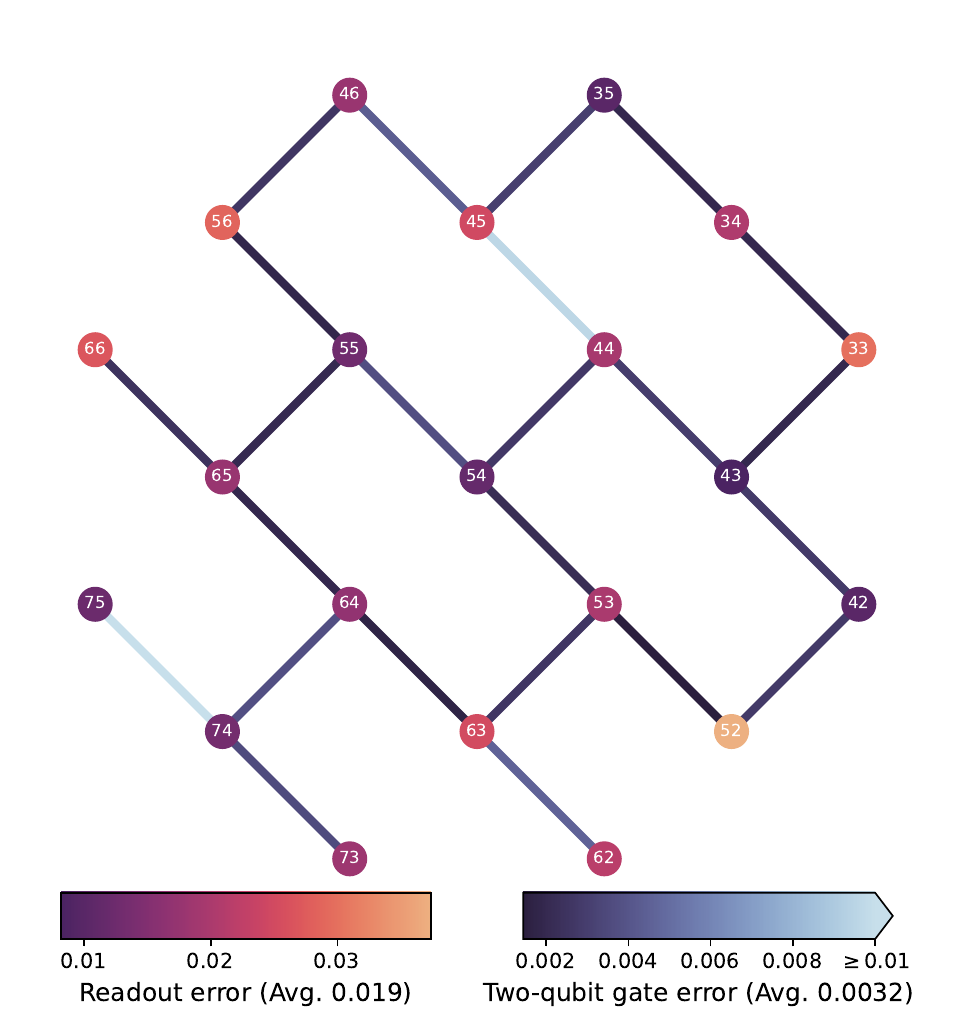}
    }
    \hfill
    \subfloat[Second distance-3 qubit configuration.\label{fig:acid_d3_x_2}]{
        \includegraphics[width=0.3\textwidth]{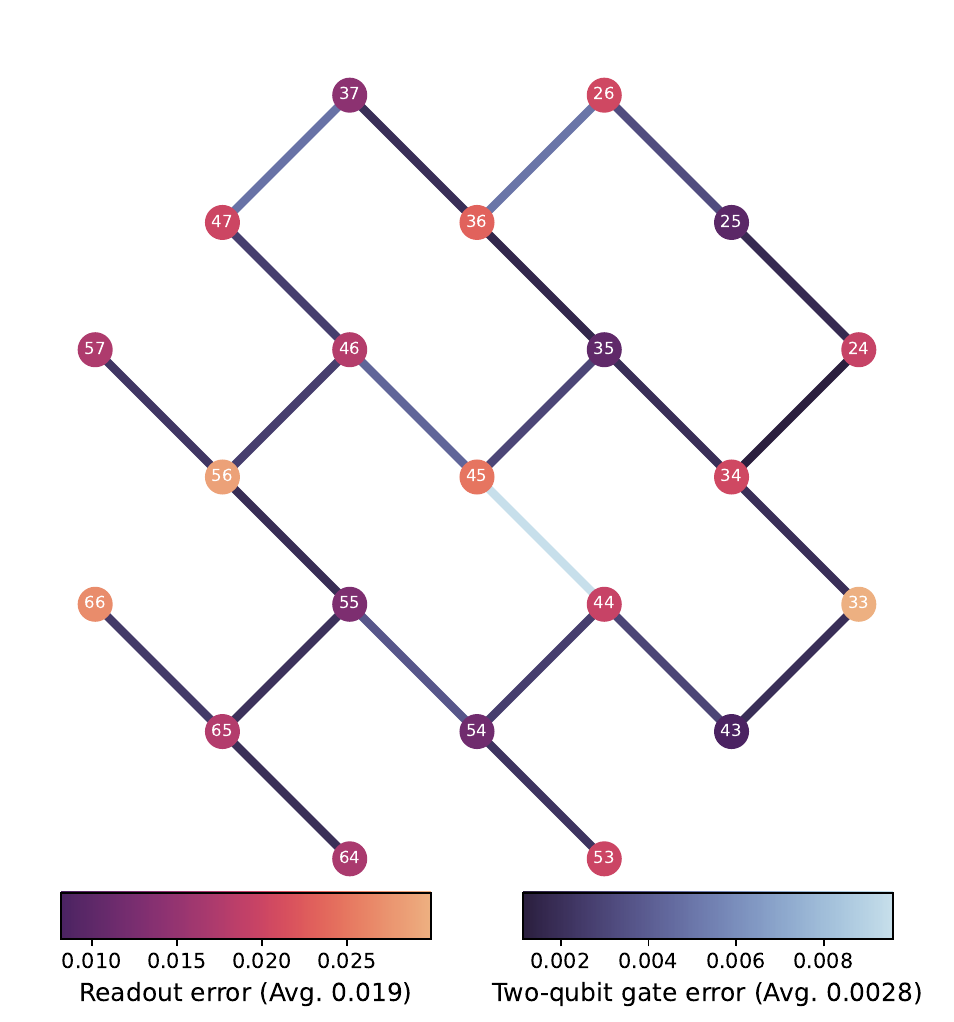}
    }
    \hfill
    \subfloat[Third distance-3 qubit configuration.\label{fig:acid_d3_x_3}]{
        \includegraphics[width=0.3\textwidth]{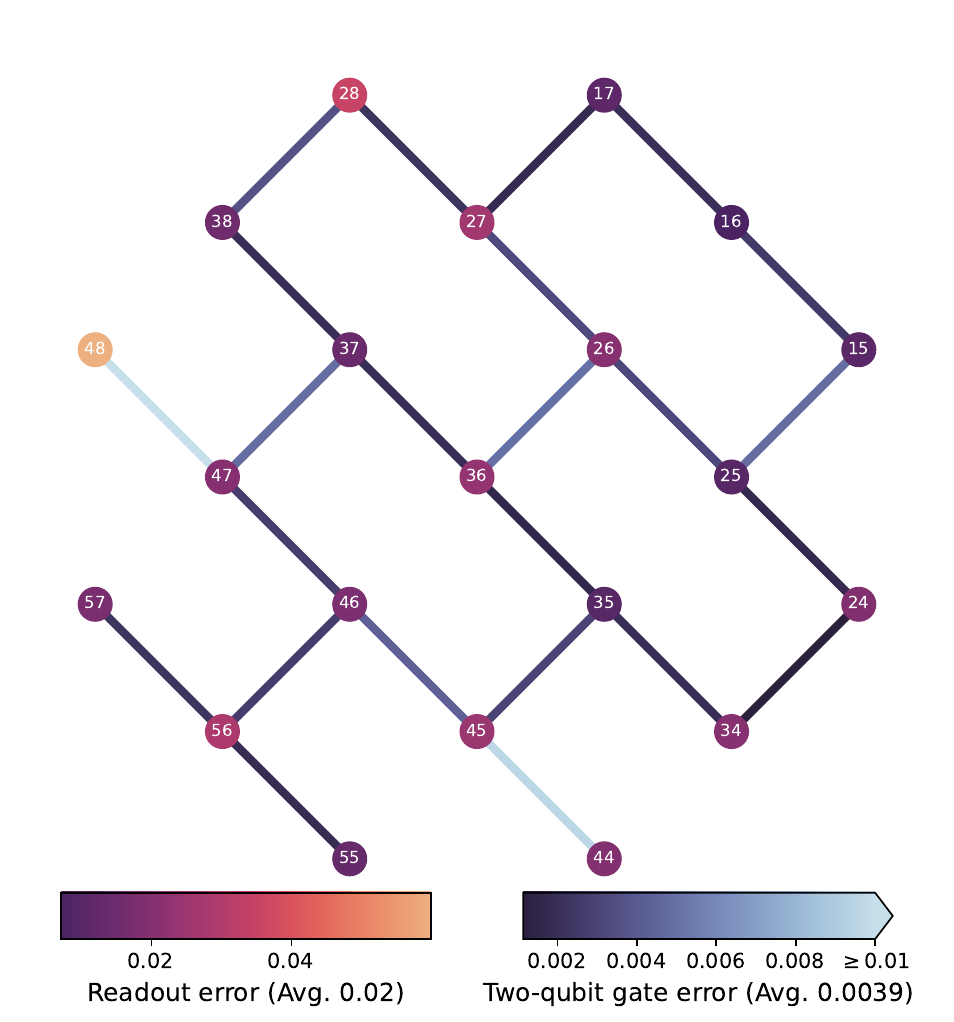}
    }
    
    \vspace{1em}
    
    \subfloat[Detection probabilities.\label{fig:acid_d3_x_prob}]{
        \includegraphics[width=\textwidth]{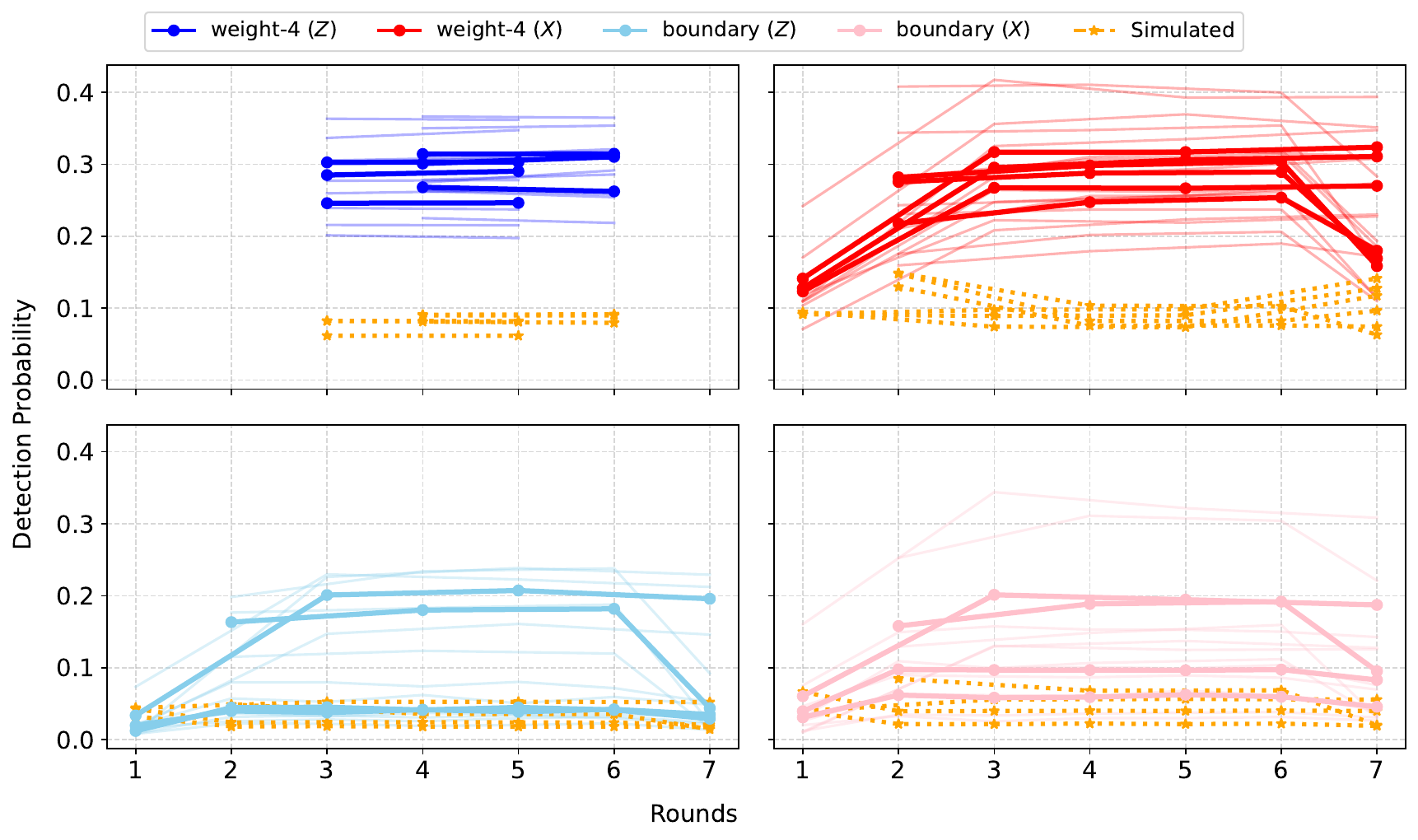}
    }
    
    \vspace{1em}
    
    \subfloat[Heatmap of detection probabilities over rounds.\label{fig:acid_d3_x_heat}]{
        \includegraphics[width=\textwidth]{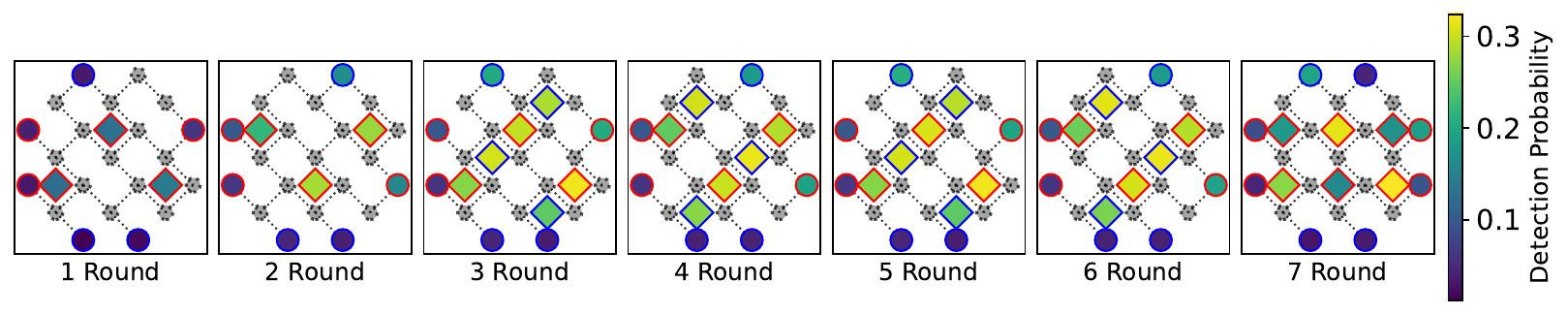}
    }
    
    \caption{Qubit configurations and reset-free detection probabilities for distance-3 LUCI baseline codes prepared and measured in the X basis. Results were obtained using $10^4$ samples per patch on the \textit{ibm\_miami} device.}
    \label{fig:LUCI_d3_X_analysis}
\end{figure*}

\begin{figure*}[h]
    \centering
    
    \subfloat[$d_X=5$ and $d_Z=3$ qubit configuration.\label{fig:acid_d53_z_config}]{
        \includegraphics[width=0.6\textwidth]{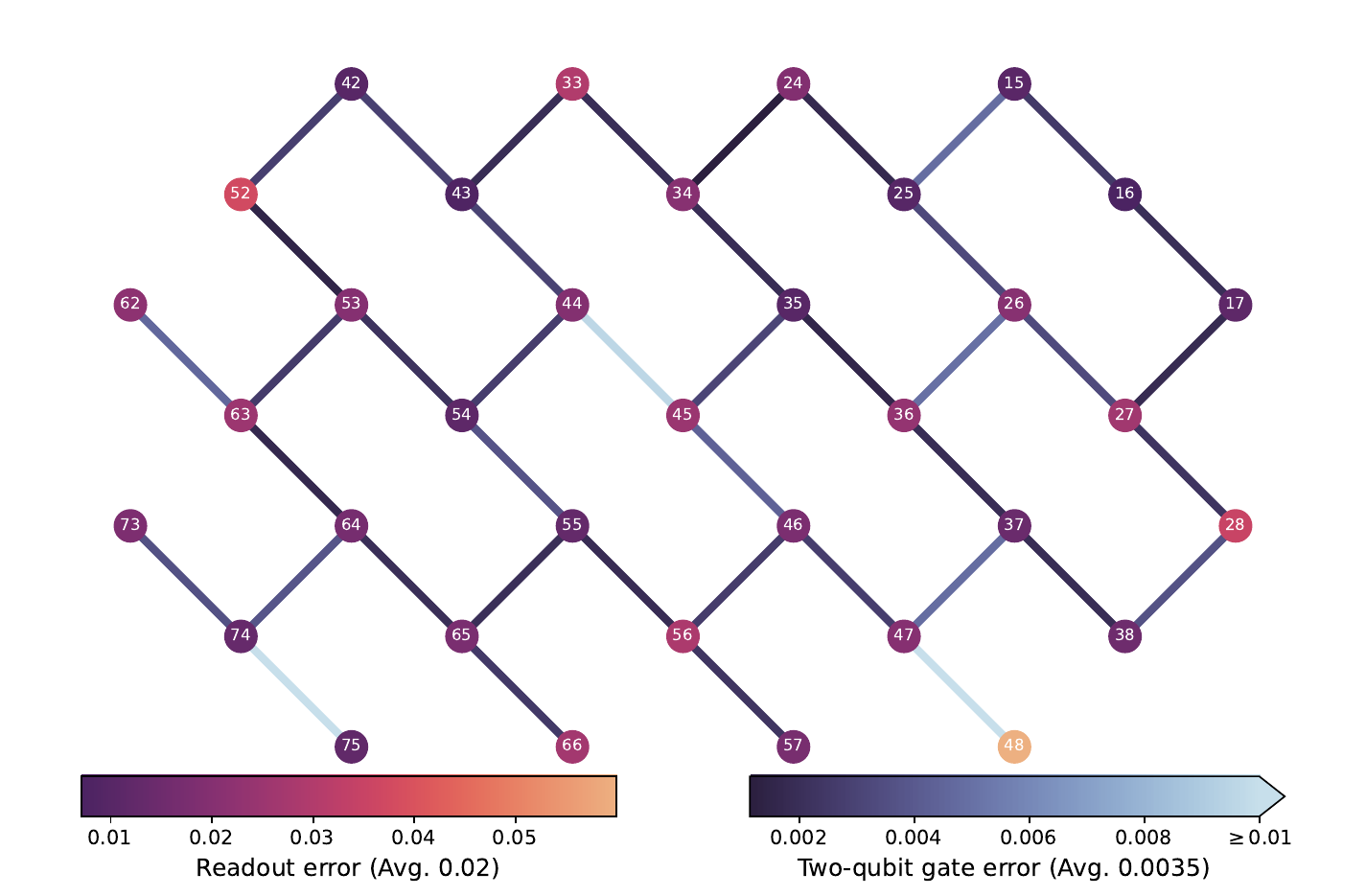}
    }
    
    \vspace{1em}
    
    \subfloat[Detection probabilities.\label{fig:acid_d53_z_prob}]{
        \includegraphics[width=0.9\textwidth]{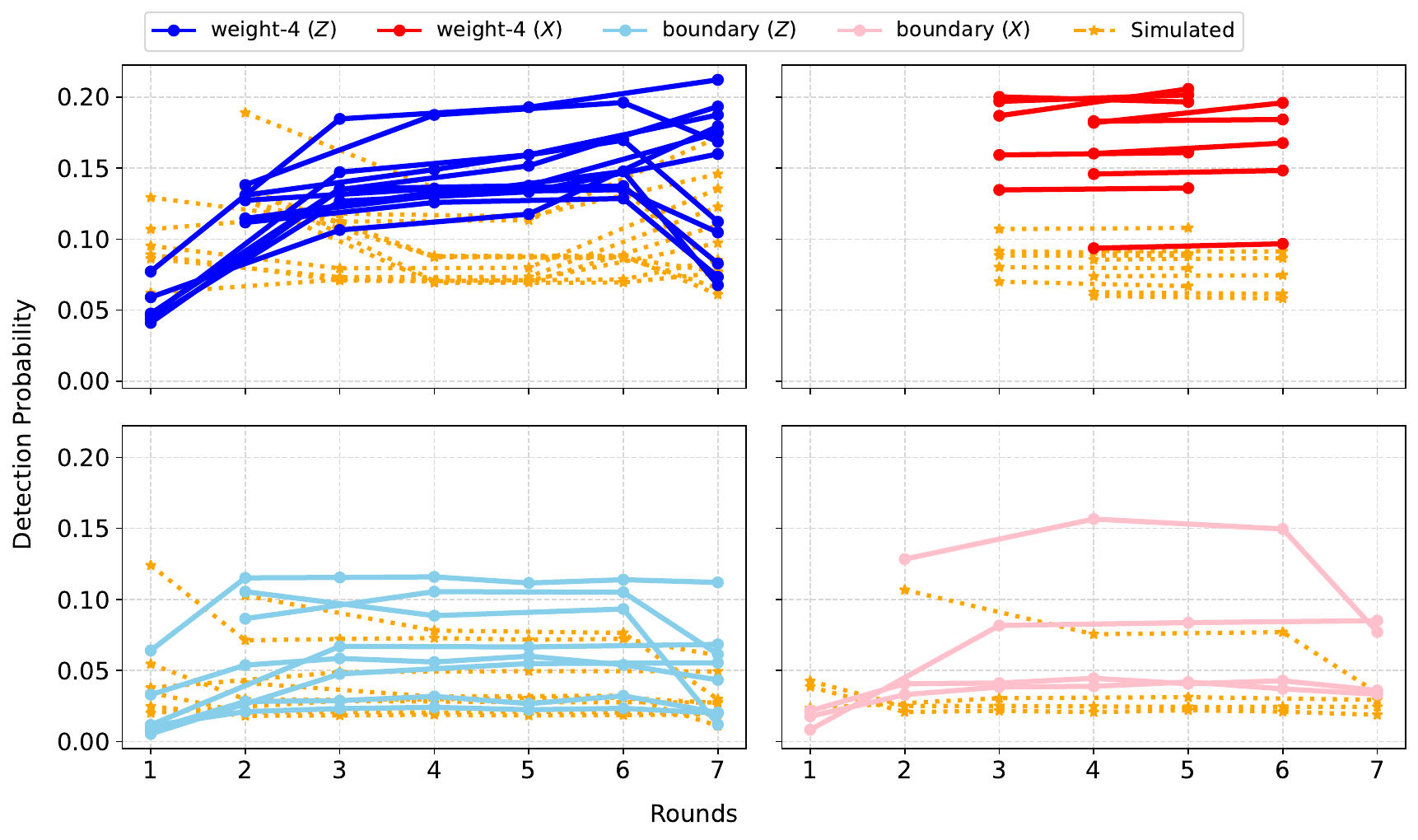}
    }
    
    \vspace{1em}
    
    \subfloat[Heatmap of detection probabilities over rounds.\label{fig:acid_d53_z_heat}]{
        \includegraphics[width=\textwidth]{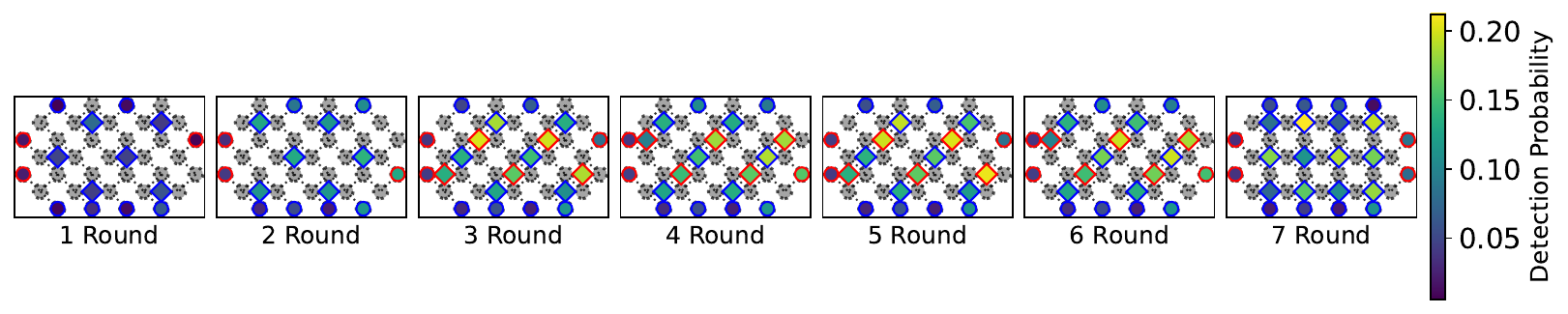}
    }
    
    \caption{Qubit configurations and reset-free detection probabilities for $d_X=5$ and $d_Z=3$ LUCI baseline code patch prepared and measured in the Z basis. Results were obtained using $10^4$ samples per patch on the \textit{ibm\_miami} device.}
    \label{fig:LUCI_code_d53_Z_analysis}
\end{figure*}

\begin{figure*}[h]
    \centering
    
    \subfloat[$d_X=3$ and $d_Z=5$ qubit configuration.\label{fig:acid_d35_x_config}]{
        \includegraphics[width=0.33\textwidth]{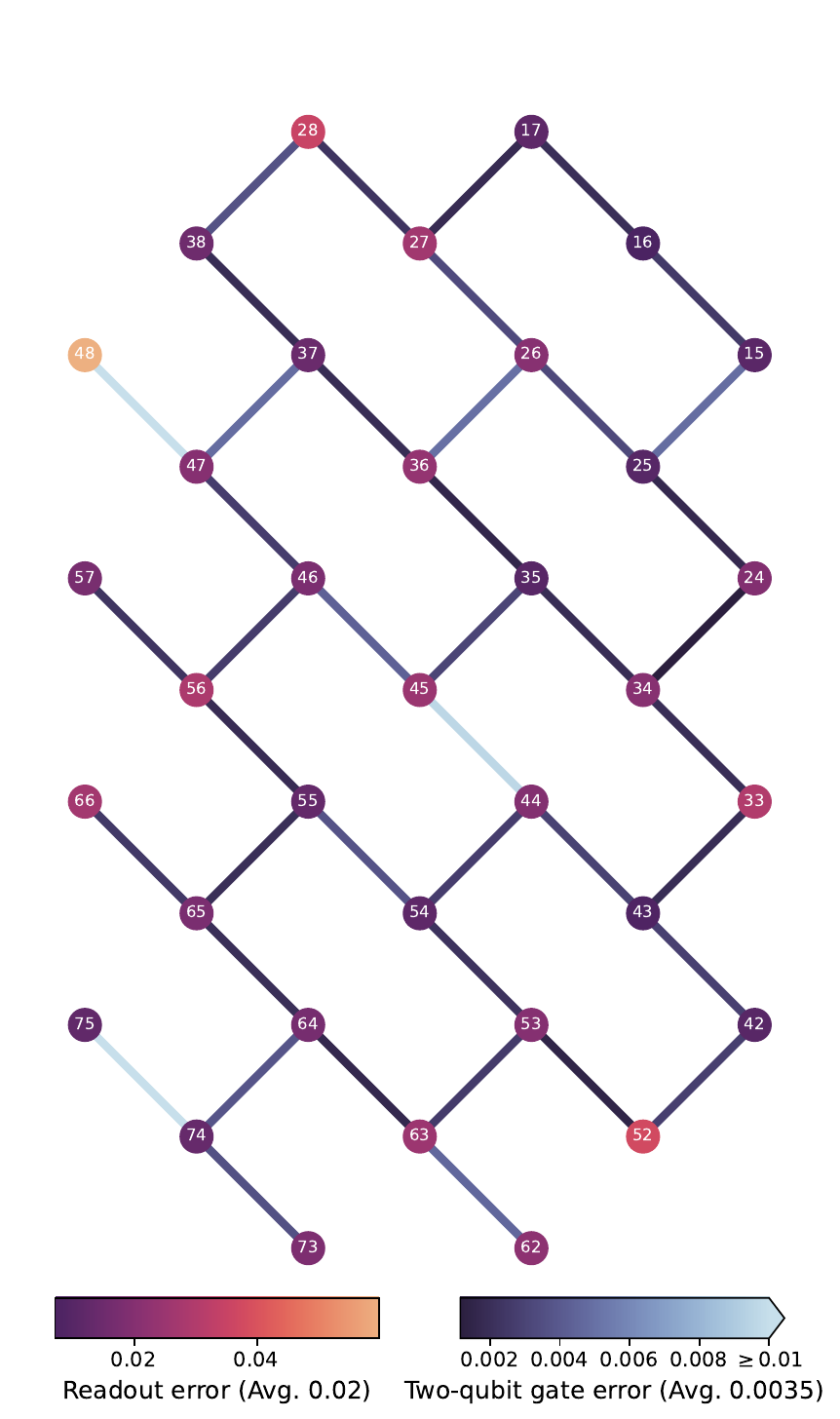}
    }
    \hfill
    \subfloat[Detection probabilities.\label{fig:acid_d35_x_prob}]{
        \includegraphics[width=0.62\textwidth]{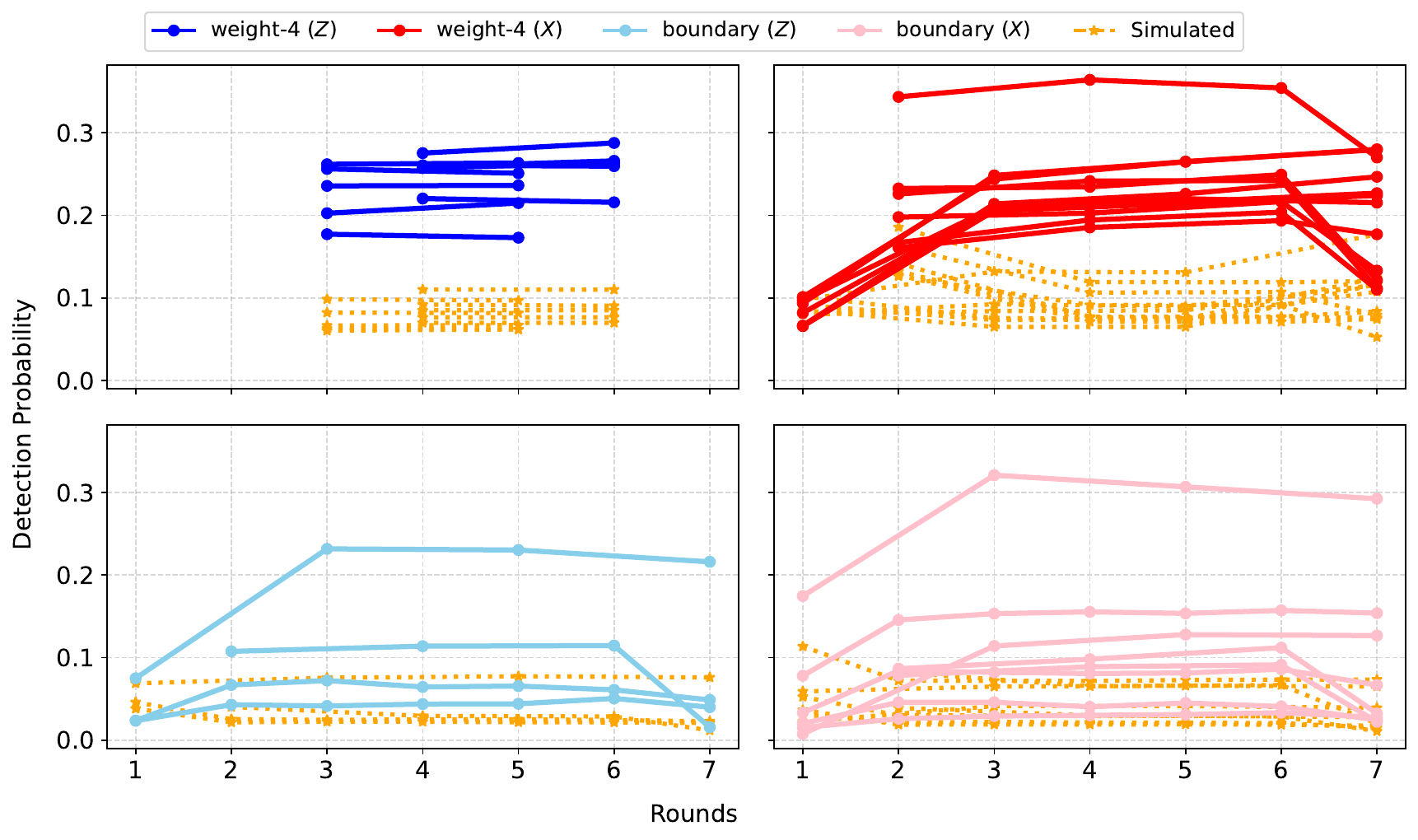}
    }
    
    \vspace{1em}
    
    \subfloat[Heatmap of detection probabilities over rounds.\label{fig:acid_d35_x_heat}]{
        \includegraphics[width=\textwidth]{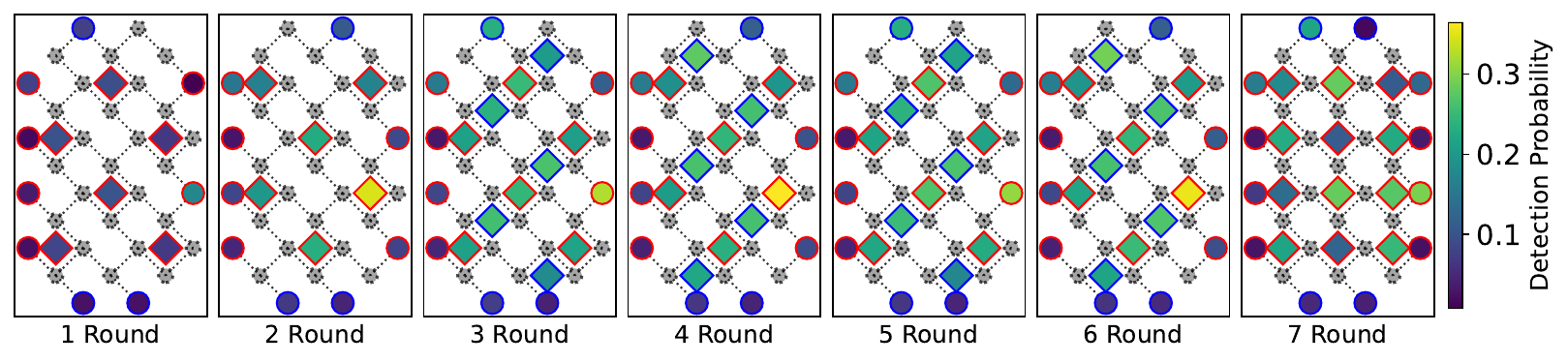}
    }
    
    \vspace{1em}
    
    \caption{Qubit configurations and reset-free detection probabilities for $d_X=3$ and $d_Z=5$ LUCI baseline code patch prepared and measured in the X basis. Results were obtained using $10^4$ samples per patch on the \textit{ibm\_miami} device.}
    \label{fig:LUCI_code_d35_X_analysis}
\end{figure*}


\begin{figure*}[h]
    \centering
    
    \subfloat[First distance-3 qubit configuration.\label{fig:var_d3_z_1}]{
        \includegraphics[width=0.3\textwidth]{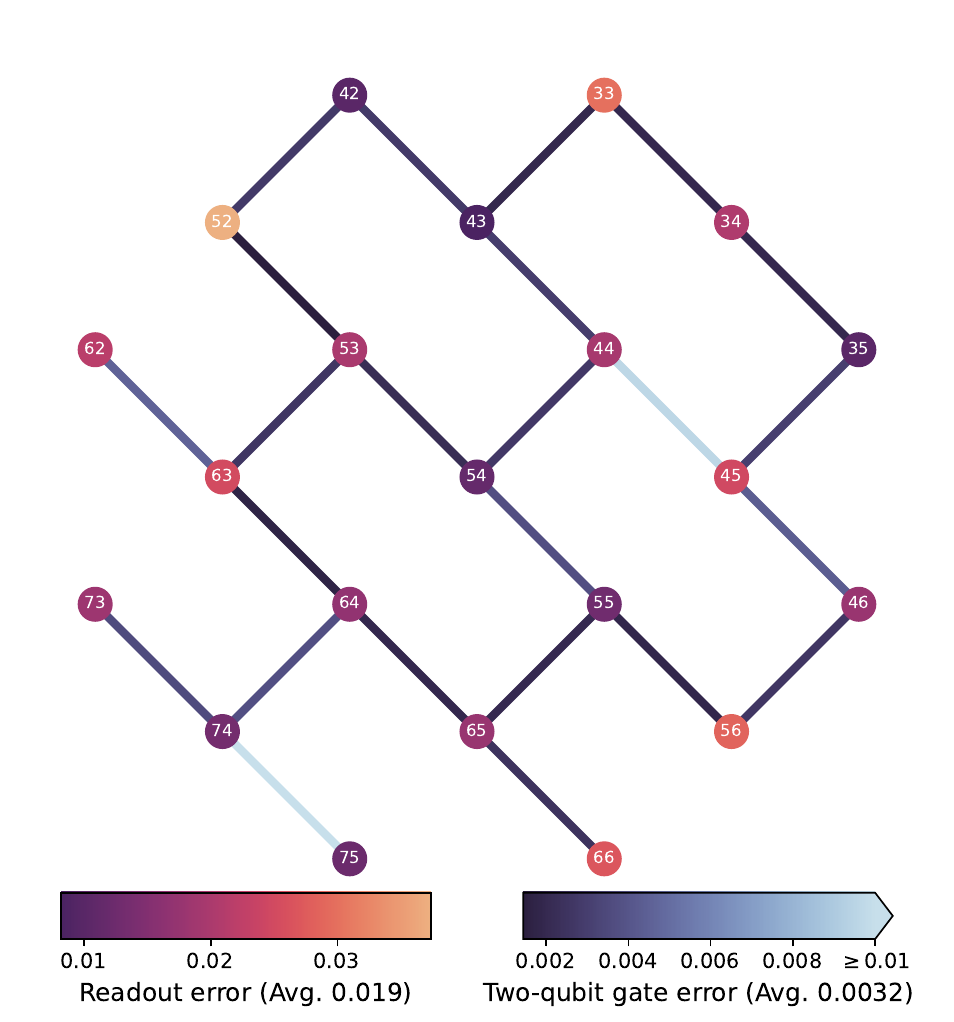}
    }
    \hfill
    \subfloat[Second distance-3 qubit configuration.\label{fig:var_d3_z_2}]{
        \includegraphics[width=0.3\textwidth]{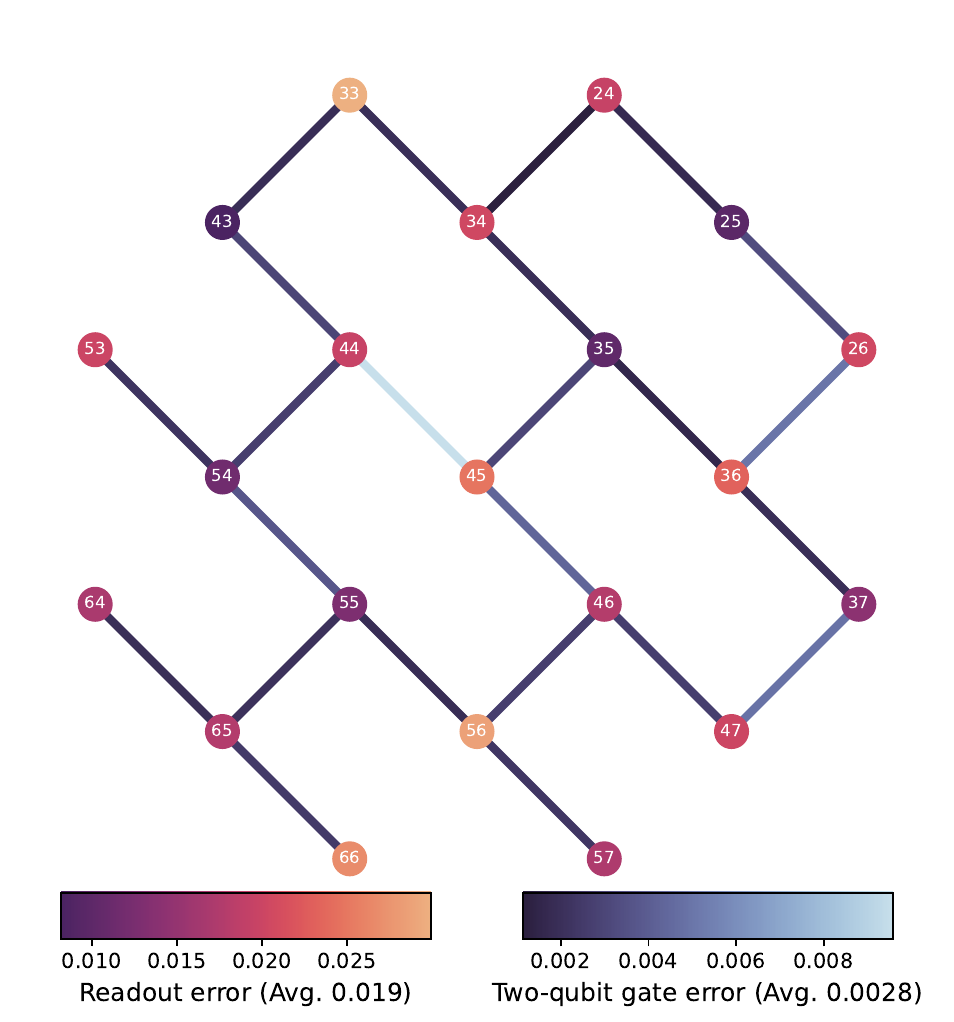}
    }
    \hfill
    \subfloat[Third distance-3 qubit configuration.\label{fig:var_d3_z_3}]{
        \includegraphics[width=0.3\textwidth]{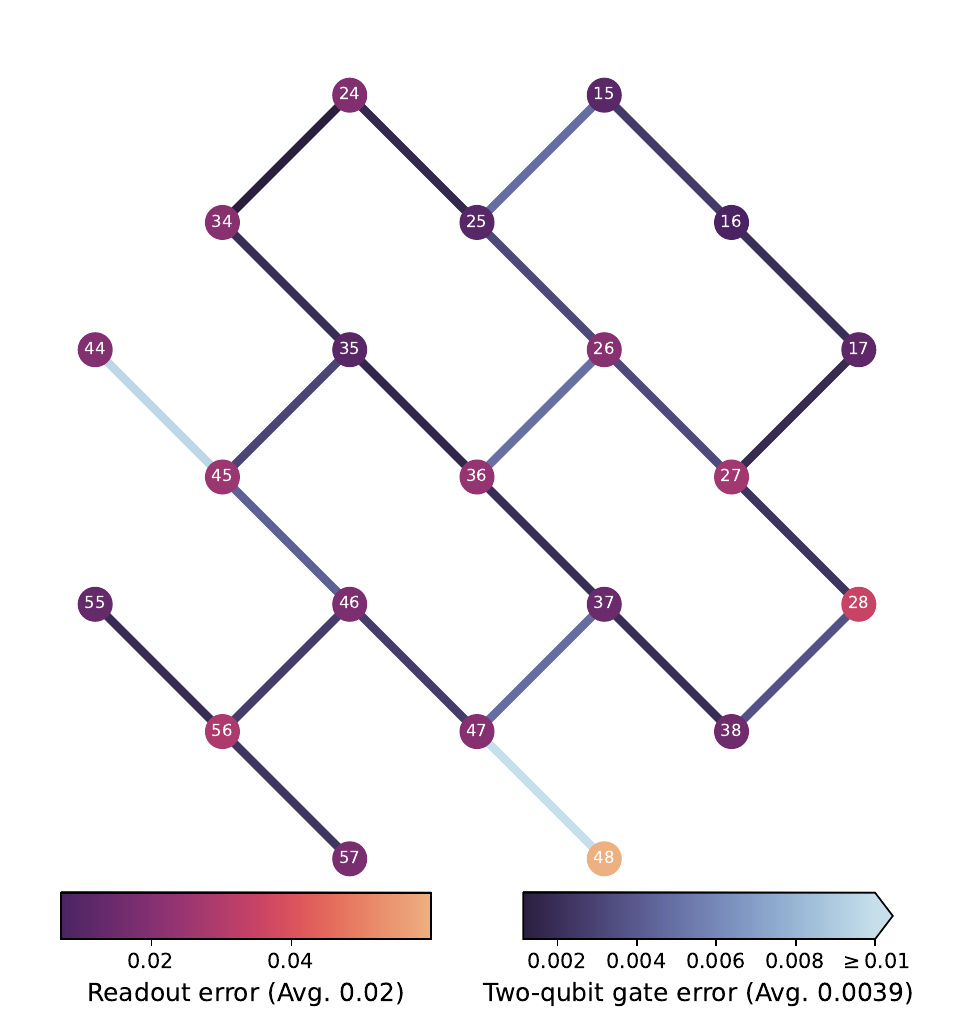}
    }
    
    \vspace{1em}
    
    \subfloat[Detection probabilities.\label{fig:var_d3_z_prob}]{
        \includegraphics[width=\textwidth]{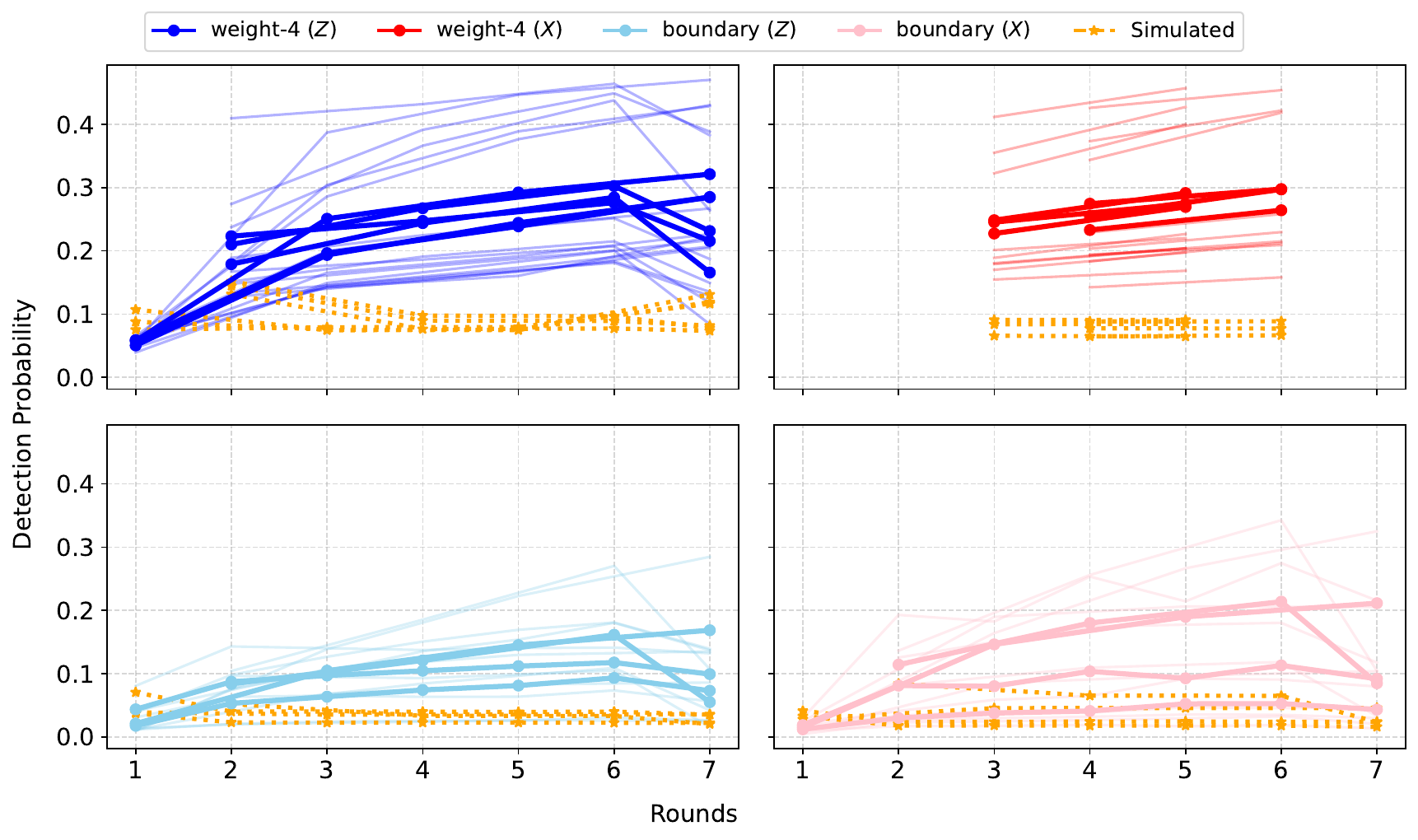}
    }

    \vspace{1em}

    \subfloat[Heatmap of detection probabilities over rounds.\label{fig:var_d3_z_heat}]{
        \includegraphics[width=\textwidth]{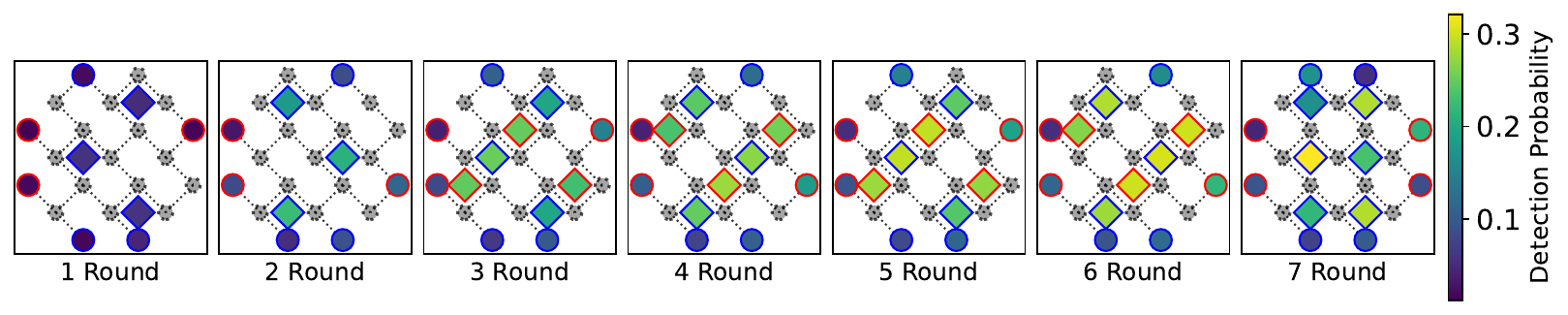}
    }
    
    \caption{Qubit configurations and reset-free detection probabilities for distance-3 LUCI variant codes prepared and measured in the Z basis. Results were obtained using $10^4$ samples per patch on the \textit{ibm\_miami} device.}
    \label{fig:LUCI_var_d3_Z_analysis}
\end{figure*}

\begin{figure*}[h]
    \centering
    
    \subfloat[First distance-3 qubit configuration.\label{fig:var_d3_x_1}]{
        \includegraphics[width=0.3\textwidth]{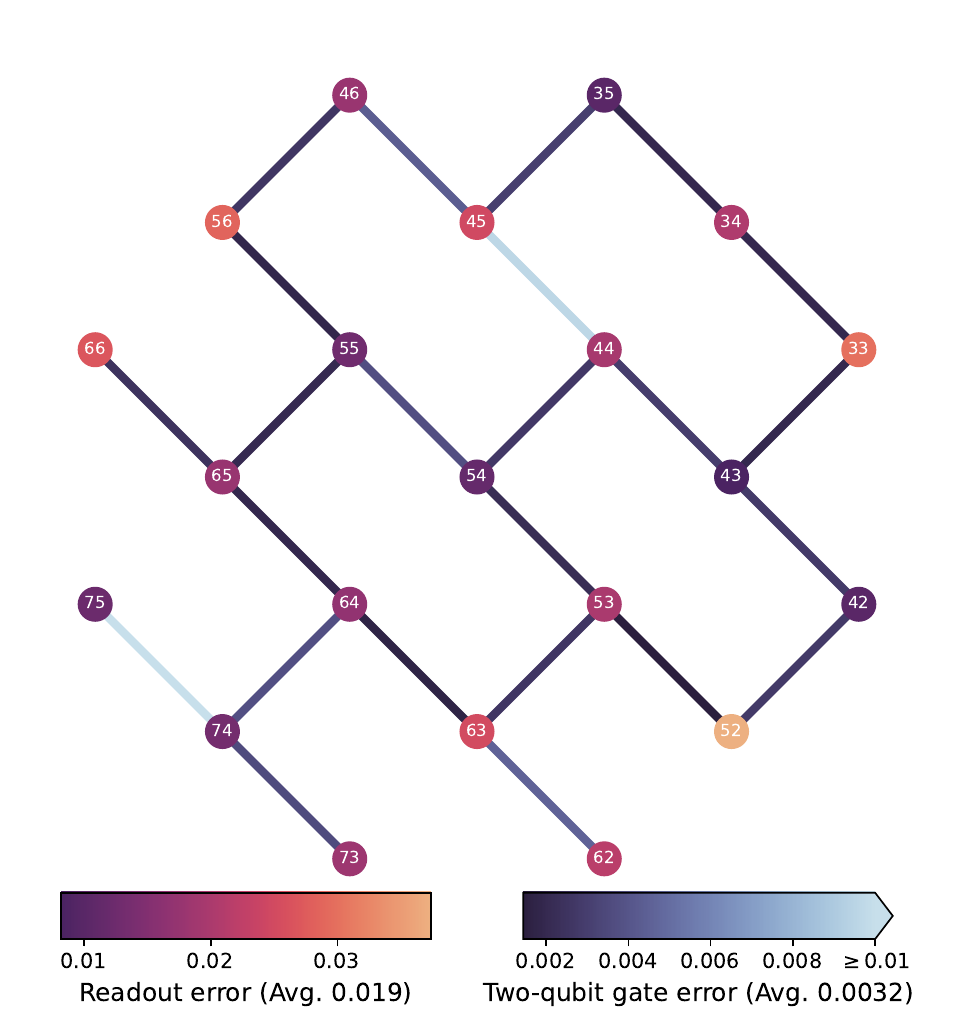}
    }
    \hfill
    \subfloat[Second distance-3 qubit configuration.\label{fig:var_d3_x_2}]{
        \includegraphics[width=0.3\textwidth]{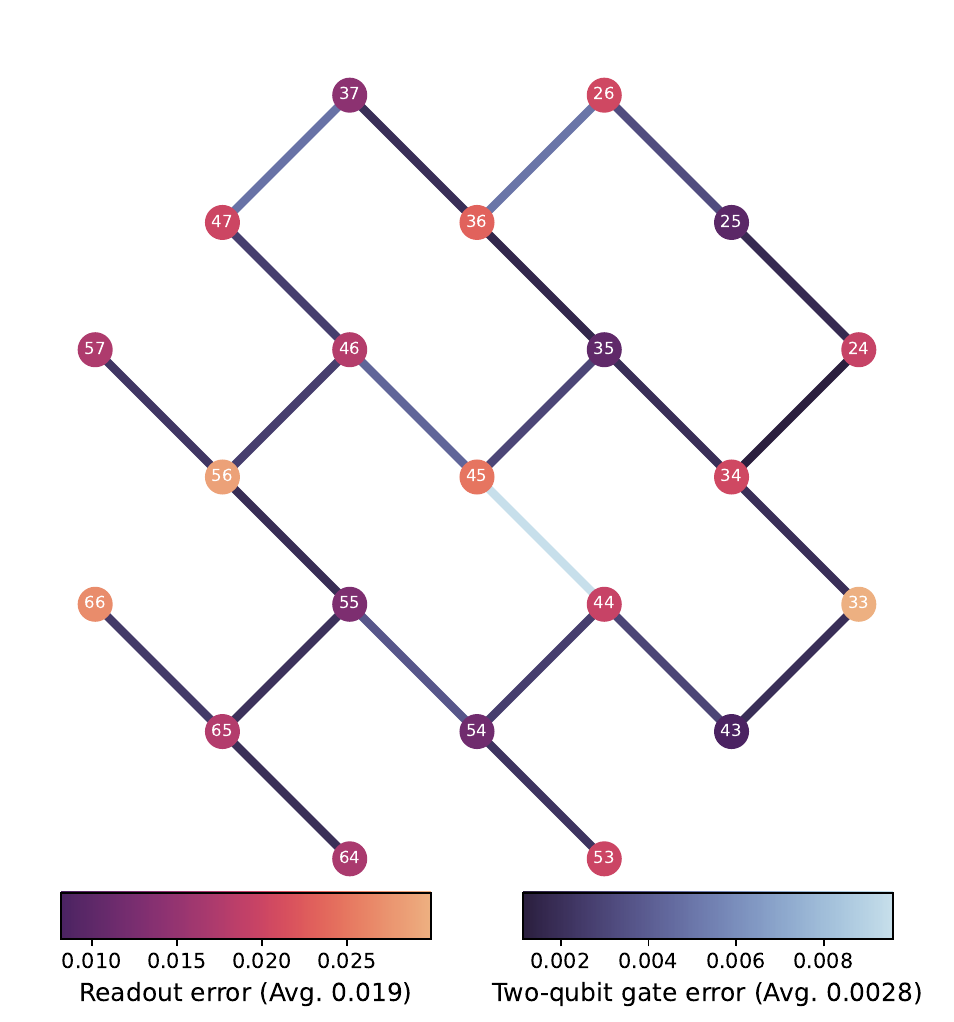}
    }
    \hfill
    \subfloat[Third distance-3 qubit configuration.\label{fig:var_d3_x_3}]{
        \includegraphics[width=0.3\textwidth]{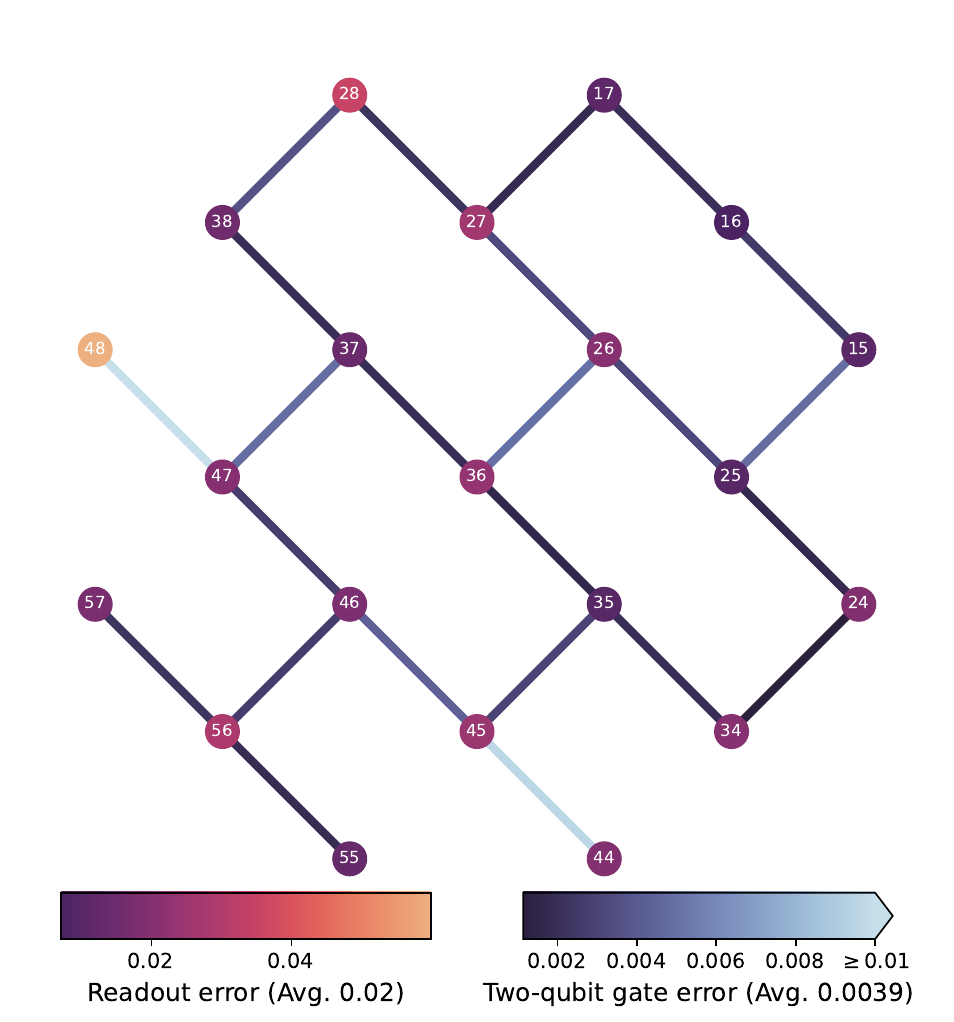}
    }
    
    \vspace{1em}
    
    \subfloat[Detection probabilities.\label{fig:var_d3_x_prob}]{
        \includegraphics[width=\textwidth]{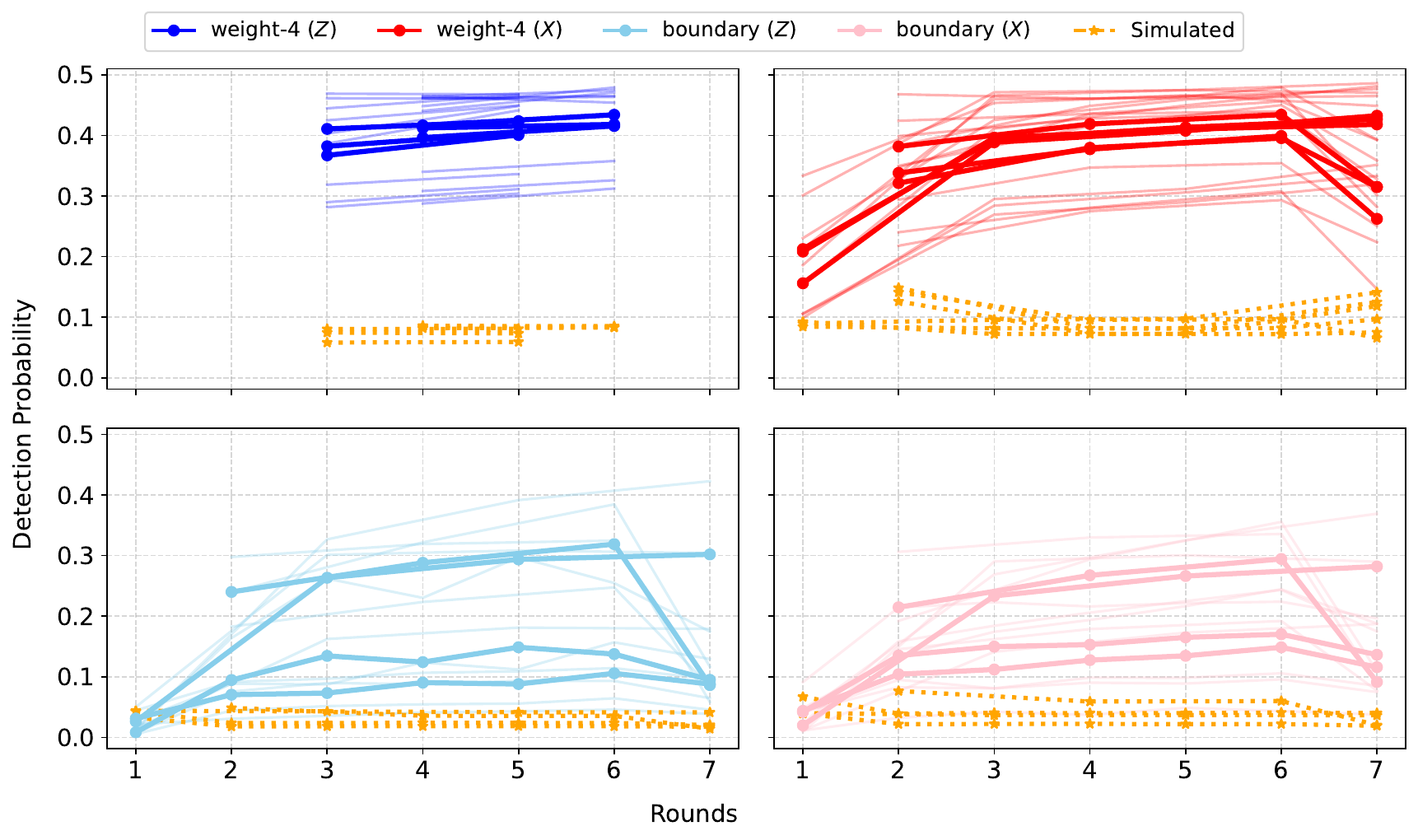}
    }

    \vspace{1em}

    \subfloat[Heatmap of detection probabilities over rounds.\label{fig:var_d3_x_heat}]{
        \includegraphics[width=\textwidth]{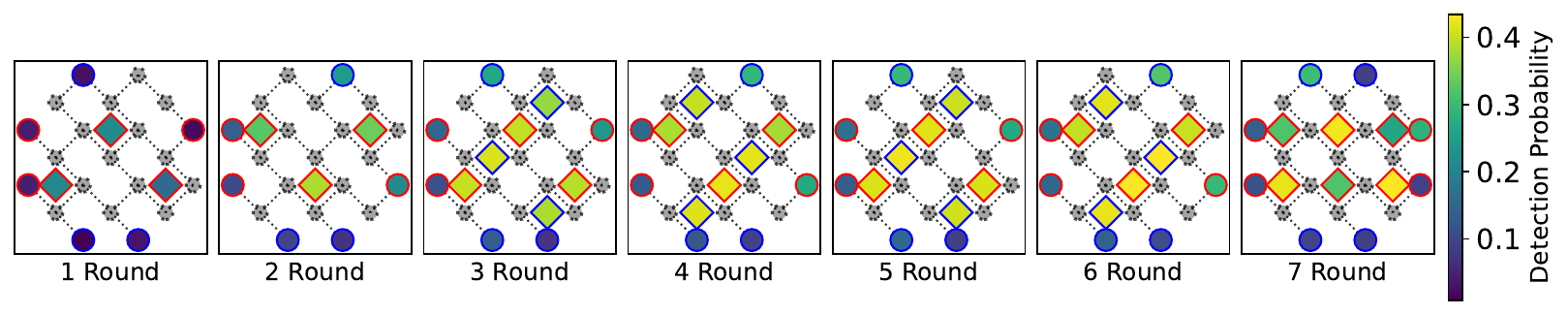}
    }
    
    \caption{Qubit configurations and reset-free detection probabilities for distance-3 LUCI variant codes prepared and measured in the X basis. Results were obtained using $10^4$ samples per patch on the \textit{ibm\_miami} device.}
    \label{fig:LUCI_var_d3_X_analysis}
\end{figure*}

\begin{figure*}[h]
    \centering
    
    \subfloat[$d_X=5$ and $d_Z=3$ qubit configuration.\label{fig:var_d53_z_config}]{
        \includegraphics[width=0.6\textwidth]{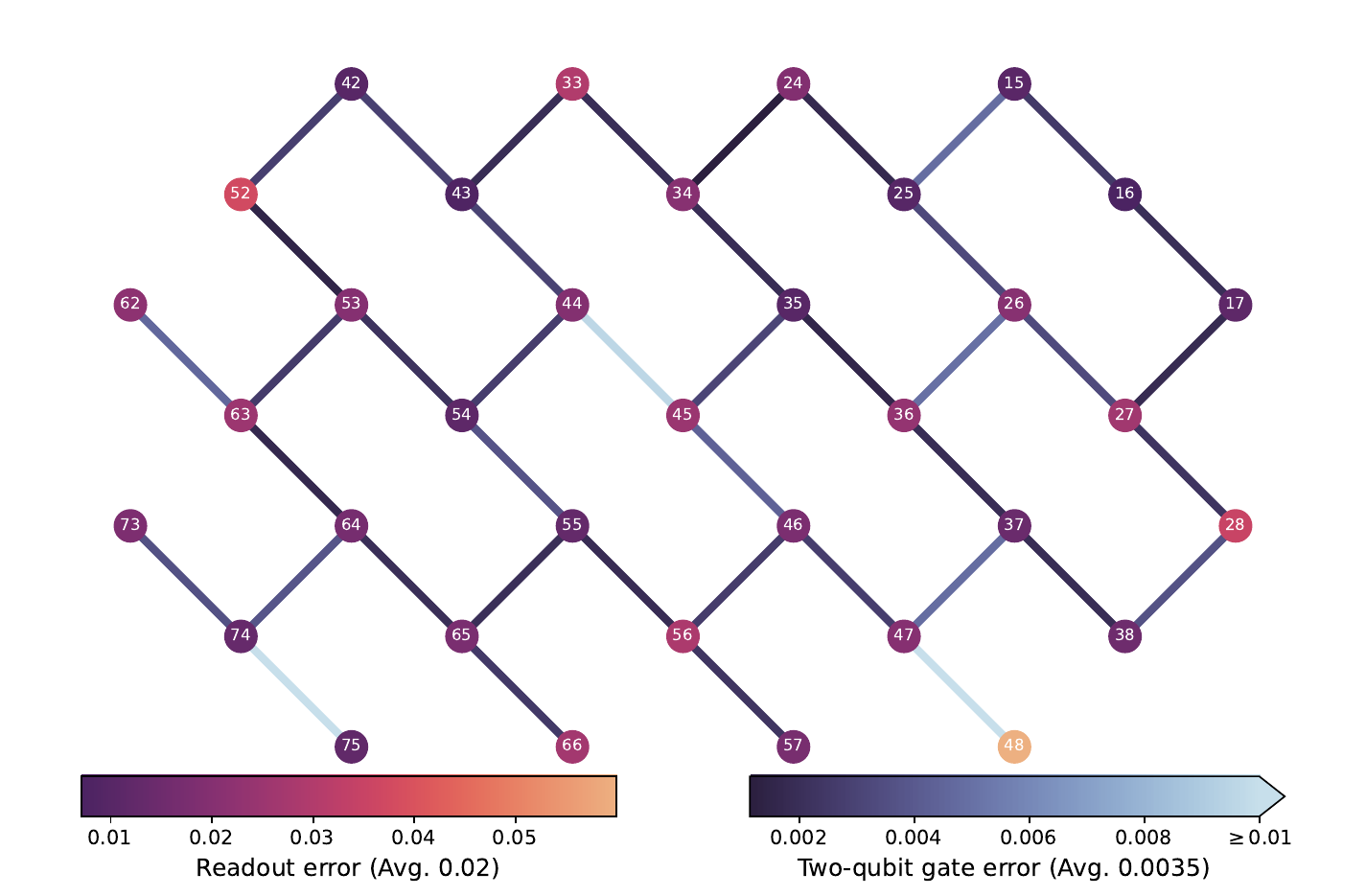}
    }
    
    \vspace{1em}
    
    \subfloat[Detection probabilities.\label{fig:var_d53_z_prob}]{
        \includegraphics[width=0.9\textwidth]{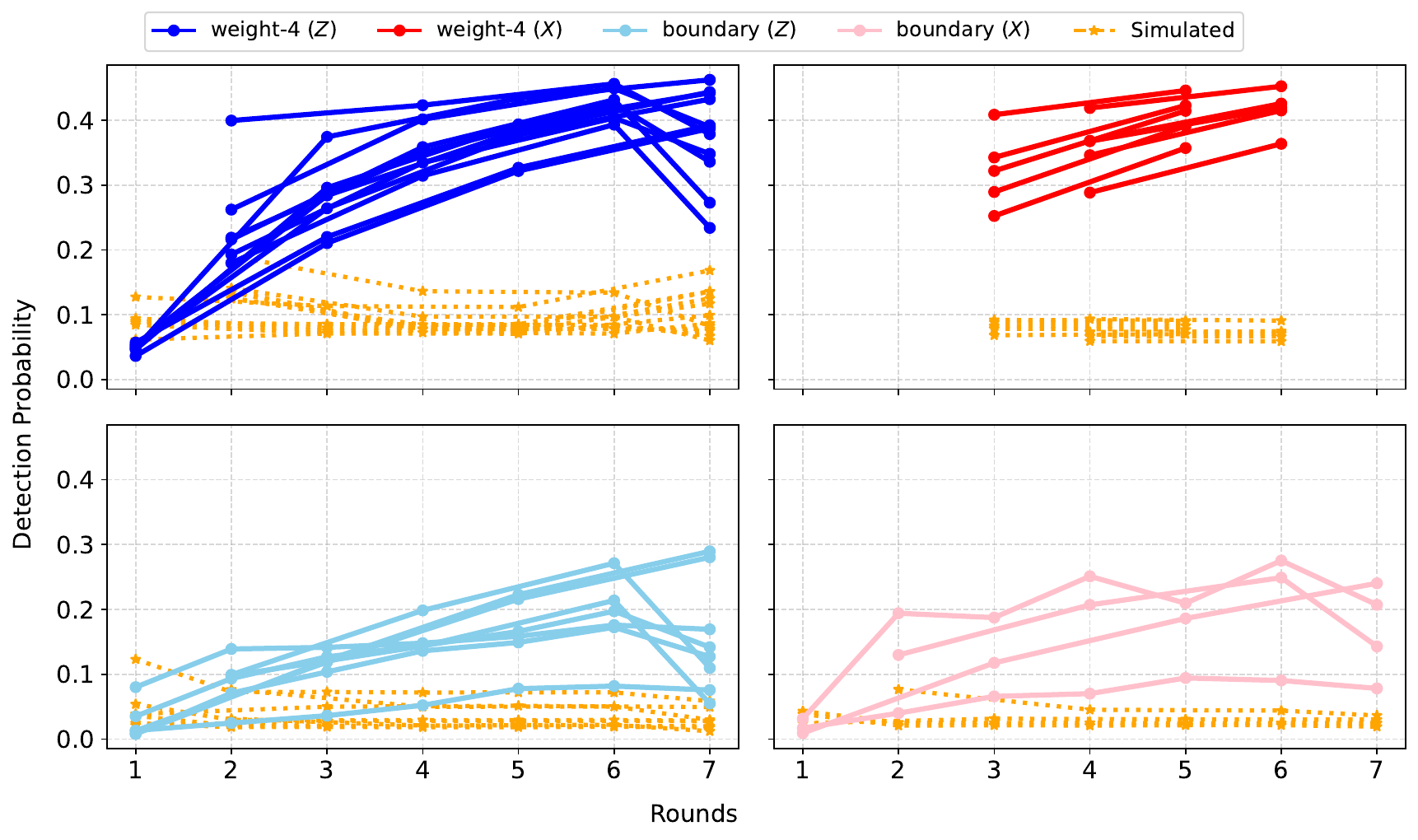}
    }
    
    \vspace{1em}
    
    \subfloat[Heatmaps of detection probabilities over rounds.\label{fig:var_d53_z_heat}]{
        \includegraphics[width=\textwidth]{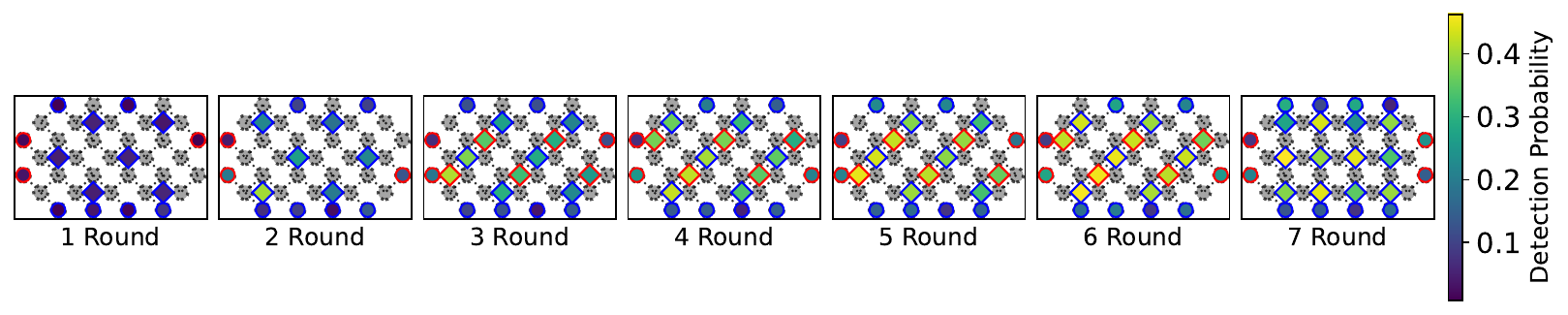}
    }
    
    \caption{Qubit configurations and reset-free detection probabilities for $d_X=5$ and $d_Z=3$ LUCI variant code patch prepared and measured in the Z basis. Results were obtained using $10^4$ samples per patch on the \textit{ibm\_miami} device.}
    \label{fig:LUCI_var_d53_Z_analysis}
\end{figure*}

\begin{figure*}[h]
    \centering
    
    \subfloat[$d_X=3$ and $d_Z=5$ qubit configuration.\label{fig:var_d35_x_config}]{
        \includegraphics[width=0.33\textwidth]{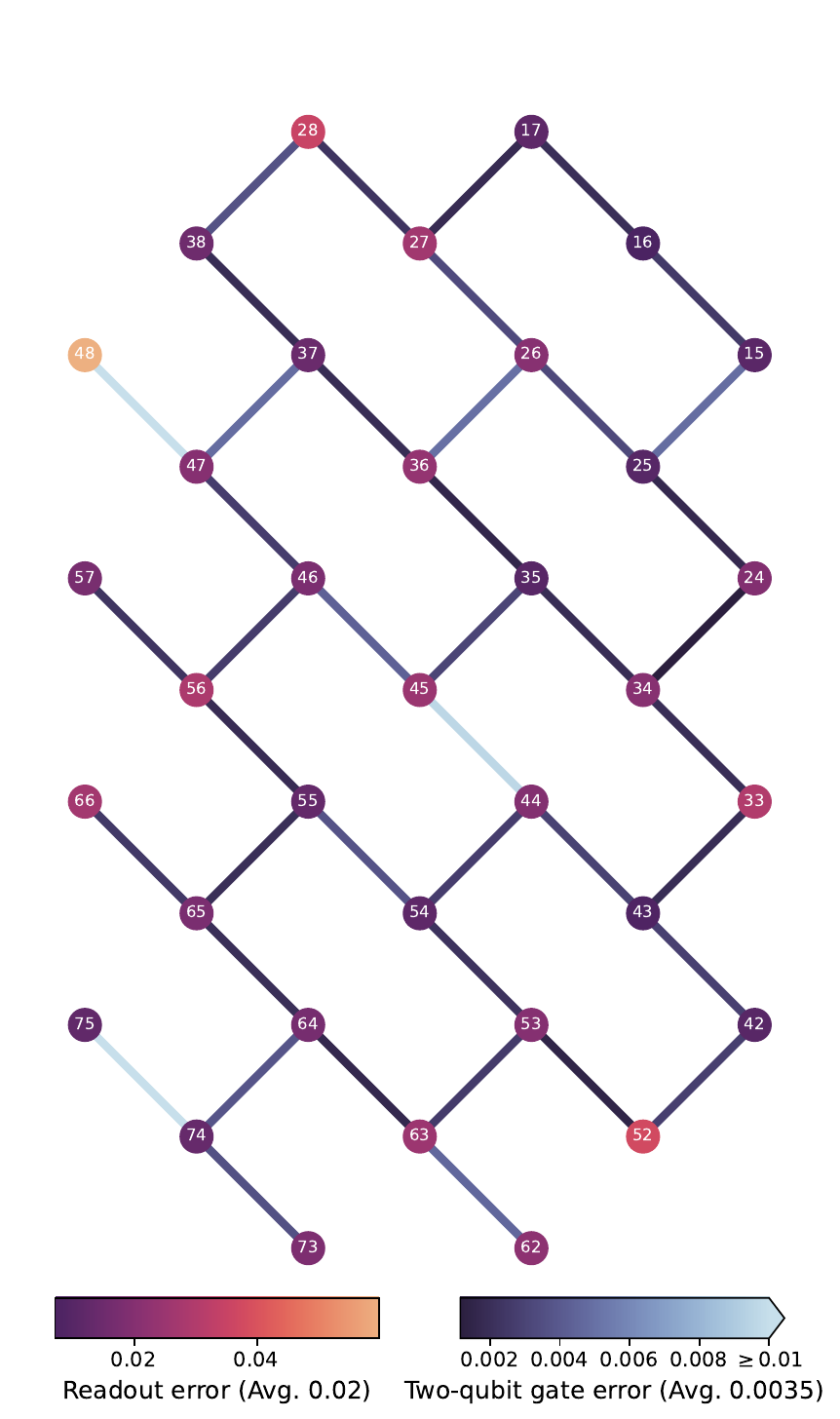}
    }
    \hfill
    \subfloat[Detection probabilities.\label{fig:var_d35_x_prob}]{
        \includegraphics[width=0.62\textwidth]{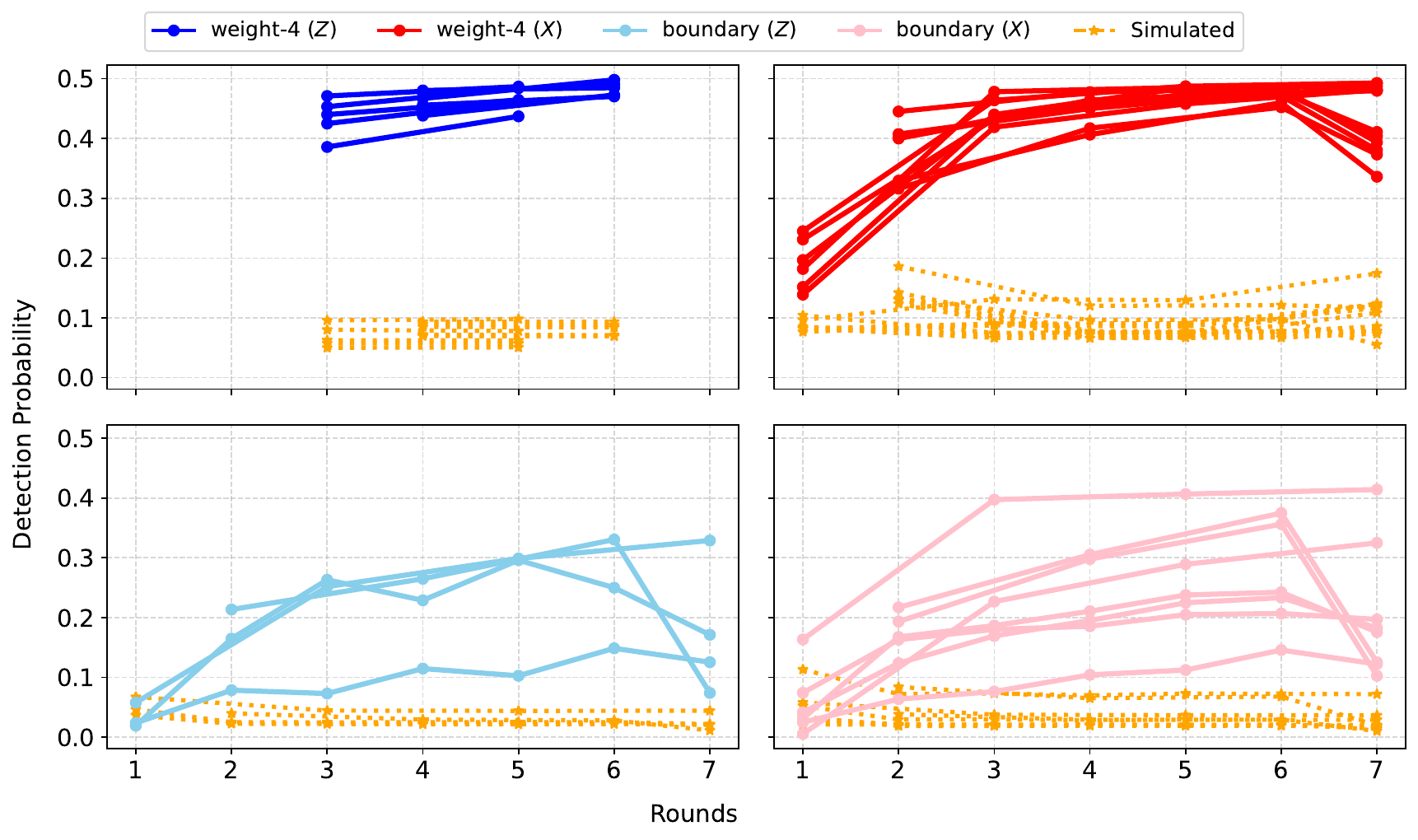}
    }
    
    \vspace{1em}
    
    \subfloat[Heatmap of detection probabilities over rounds.\label{fig:var_d35_x_heat}]{
        \includegraphics[width=\textwidth]{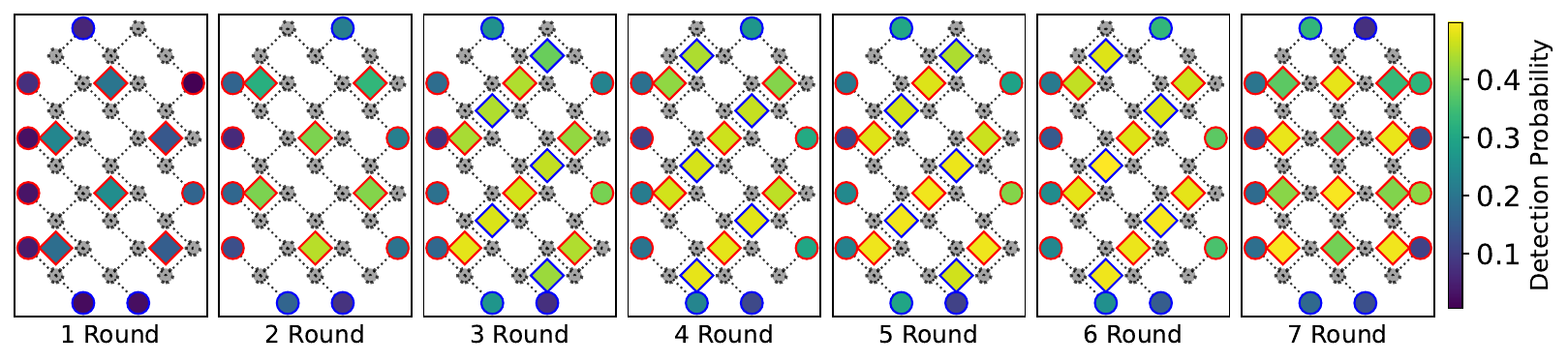}
    }
    
    \vspace{1em}
    
    \caption{Qubit configurations and reset-free detection probabilities for $d_X=3$ and $d_Z=5$ LUCI variant code patch prepared and measured in the X basis. Results were obtained using $10^4$ samples per patch on the \textit{ibm\_miami} device.}
    \label{fig:LUCI_var_d35_X_analysis}
\end{figure*}

\end{document}